\documentclass[
 reprint,
superscriptaddress,
 amsmath,amssymb,
 aps,
 prapplied,
]{revtex4-2}

\usepackage{graphicx}
\usepackage{dcolumn}
\usepackage{bm}
\usepackage[utf8]{inputenc}
\usepackage[T1]{fontenc}
\usepackage{mathptmx}
\usepackage[caption=false,font=footnotesize]{subfig}
\usepackage{tabularx}
\usepackage{floatrow}
\usepackage{textgreek}
\usepackage{placeins}

\begin{document}

\preprint{APS/123-QED}




\title{Bandwidth Limit and Synthesis Approach for Single Resonance Ultrathin Metasurfaces}


 
\author{Ashif A. Fathnan}
\affiliation{School of Engineering and Information Technology, \\
University of New South Wales (UNSW) Canberra, Australia}
\email{a.fathnan@student.unsw.edu.au}

\author{Andreas E. Olk}%
\affiliation{School of Engineering and Information Technology, \\
University of New South Wales (UNSW) Canberra, Australia}
\affiliation{IEE S.A., 1, rue de Campus, 7795 Bissen, Luxembourg}%

\author{David A. Powell}%
\affiliation{School of Engineering and Information Technology, \\
University of New South Wales (UNSW) Canberra, Australia}

\date{\today}

\begin{abstract}
Metasurfaces have emerged as a promising technology for the manipulation of electromagnetic waves within a thin layer. 
In planar ultrathin metasurfaces, there exist rigorous narrowband design methods, based on the equivalent surface impedance of patterned metallic layers on dielectric substrates. In this work, we derive a limit on bandwidth achievable in these metasurfaces, based on constraints that their meta-atoms should be passive, causal and lossless, and that they should obey the time-bandwidth product rules of a single resonance structure. The results show that in addition to elementary design parameters involving variation of the surface impedance, the bandwidth is critically limited by the dielectric substrate thickness and permittivity.  
We then propose a synthesis method for broadband ultrathin metasurfaces, based on an LC resonance fit of the required surface impedance, and experimentally verify a broadband dispersive structure at millimeter-wave frequencies. This results in a bandwidth enhancement  of over 90\%, relative to a reference metasurface created with the narrowband design process. 
\end{abstract}

\maketitle

\section{Introduction} \label{sec:introduction}

Metasurfaces are a unique type of metamaterial, where control of propagating waves can be achieved with thin layers of scatterering elements. A common metasurface architecture is based on patterned metallic layers, where the use of equivalent surface impedances enables robust design methods which can realize essentially arbitrary wavefront manipulation functions \cite{diaz2017generalized,pfeiffer2013millimeter,ra2014tailoring,asadchy2016perfect}. Utilizing this architecture, a single resonant response is typically exploited to introduce abrupt phase changes of the radiating wave over the full $2\pi$ range. By detuning individual elements (known as meta-atoms) to engineer their phase shift (or equivalently, their surface impedance), efficient wavefront control can be achieved, but it is often limited to a narrow bandwidth around the design frequency. 

Techniques to overcome the bandwidth limitations of metasurfaces can be divided into broadband dispersive and achromatic approaches.  In the broadband dispersive approach, the reflection or refraction angle is allowed to vary with frequency, while the specular reflection and unwanted diffraction orders are suppressed over a significant bandwidth \cite{gao2018vertically,pors2013broadband,sun2012high,YouTerahertzreflectarrayenhanced2019,callewaert2018inverse}. In the achromatic approach, a fixed reflection or refraction angle is maintained over the operating bandwidth by engineering the group delay spatially across the metasurface \cite{arbabi2017controlling,wang2017broadband,khorasaninejad2017achromatic,shrestha2018broadband,cheng2019broadband}. To achieve large variation of phase delay or group delay, optical metasurfaces often utilize pillars of dielectric with thickness comparable to the operating wavelength \cite{shi2018single,wang2018broadband,sawant2019mitigating}. These structures can be understood as short sections of waveguide, where significant delay can be achieved by sacrificing the requirement for sub-wavelength thickness. For infrared and optical wavelengths, the resulting dimensions are compatible with fabrication technology, however, at lower frequency ranges the required structures would be heavy, complex to fabricate, and difficult to integrate with electronic components. Therefore, the use of patterned metallic structures is more favorable for microwave, millimeter-wave and terahertz regimes. 

\begin{figure*}[t]
\includegraphics[width=0.75\linewidth]{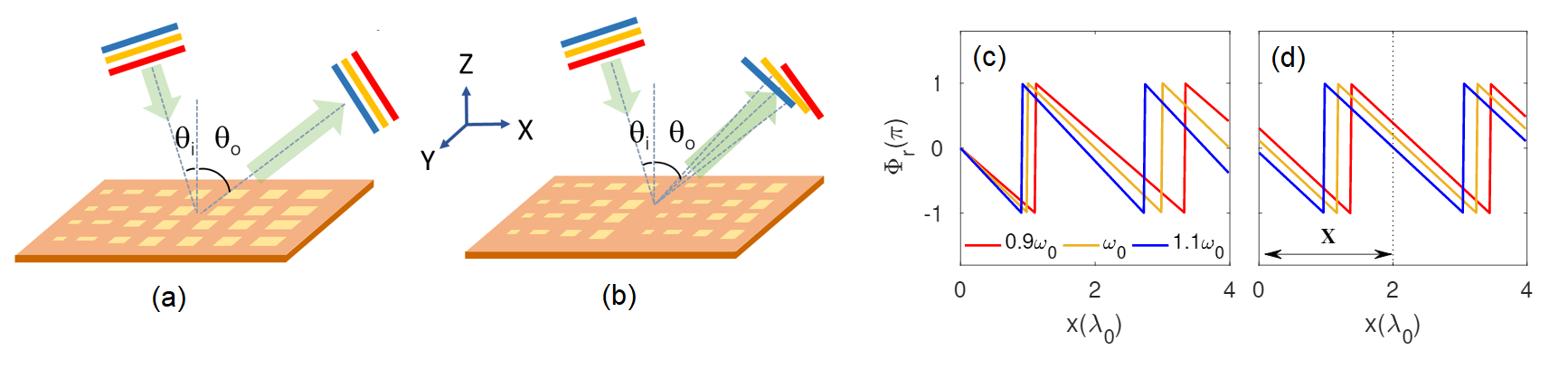}
\includegraphics[width=0.24\linewidth]{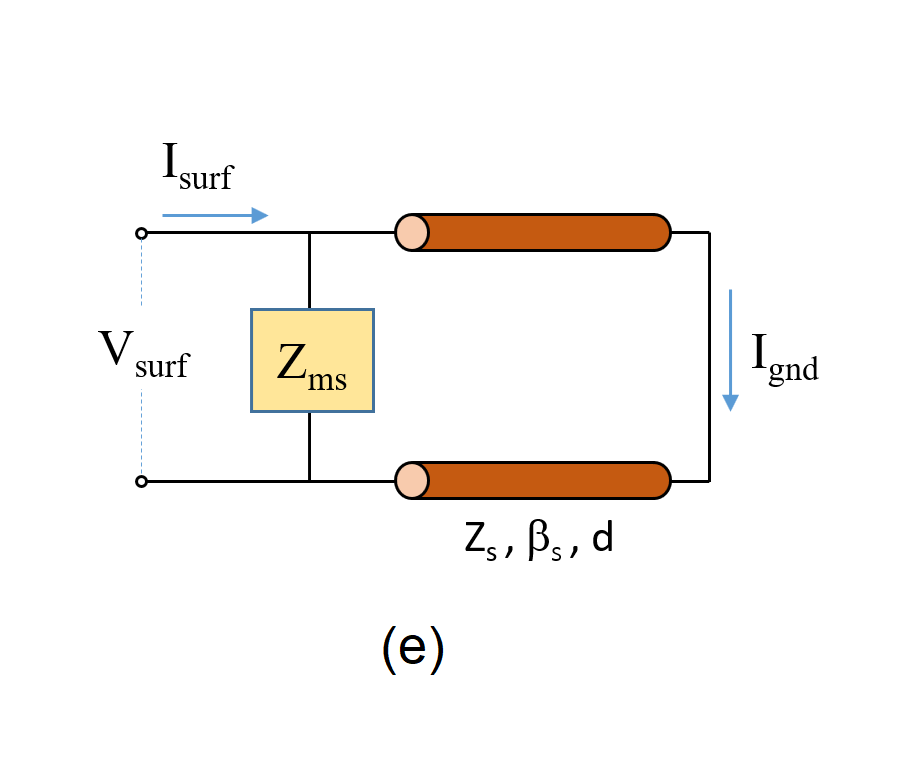}
\caption{\label{fig:ms_illustration} Metasurface performing anomalous reflection under different scenarios: (a) broadband achromatic design (constant angle responses), (b)  broadband dispersive design (frequency dependent angle). The phase requirements are shown in (c) for broadband achromatic design and (d) for broadband dispersive design. (e) Equivalent circuit of the meta-atom.}
\end{figure*}

Strategies to increase the bandwidth of metallic metasurfaces with planar architectures include the use of multi-resonance structures and thicker dielectric substrates. Multi-resonance structures have been studied in various forms, including a combination of dipoles and rings \cite{chen2013x,qu2015controlling}, multiple dipoles in parallel \cite{yoon2015broadband}, and stub-loaded dipoles \cite{YouTerahertzreflectarrayenhanced2019}. Most of these multi-resonance structures are designed using ad-hoc methods of simulation and optimization. Another approach to increase the bandwidth of thin metallic metasurfaces is by increasing the substrate thickness. In metasurfaces with metallic layers and dielectrics, it is recognized that larger dielectric thickness contributes to a lower radiative Q-factor and can therefore enhance the operational bandwidth \cite{pu2011design,zhao2013tailoring,tretyakov2015metasurfaces,hussain2017design}. However, it has been shown that when the substrate thickness is increased beyond a quarter wavelength, magnetic coupling starts to reduce, and so does the achievable bandwidth \cite{pu2013broadband}. The relationship between thickness and bandwidth has been established for structures such as thin absorbers \cite{rozanov2000ultimate} and high impedance surfaces \cite{gustafsson2011physical,brewitt2007limitation}.

For gradient printed circuit metasurfaces, we previously established that the aperture size and refraction angle limit the bandwidth \cite{fathnan2018bandwidth}. This work was based on the requirement that the impedance functions must obey Foster's reactance theorem, however the finite thickness of the substrate was neglected. In Ref.~\onlinecite{presutti2020focusing} more general bandwidth limits were introduced, based on the time-bandwidth product. Expressions were derived for different classes of metasurface, concentrating on those most relevant for optical wavelengths, such as optically thick dielectric wave-guiding structures. A single resonance limit was derived, which is more relevant for ultrathin structures, however it did not consider the retardation required to achieve a magnetic resonances, thus it did not show the influence of the substrate thickness.
Therefore, it remains unclear how refraction angle, aperture size and substrate thickness interact with each other in determining the upper limit on the bandwidth of ultrathin metasurfaces, and whether there is an optimal choice of substrate thickness.


In this paper we derive a bandwidth limit for a realistic single resonance metasurfaces, starting from a surface impedance model of printed circuit metasurfaces. In addition to the limit imposed by the reflection angle and aperture size, we quantify the critical role of substrate thickness. We show that although multi-resonance structures can increase the phase coverage, simpler geometries such as dipoles or nano-bricks can achieve similar bandwidth by choosing an optimal combination of substrate parameters. We propose a systematic method to design broadband metasurfaces, based on fitting LC resonances to the required impedance functions. 
We show that broadband anomalous reflection can be obtained by simple dog-bone and inverse dog-bone structures. We verify the design and synthesis procedure for a dispersive metasurface demonstrating anomalous reflection within the W-band (75-110\,GHz). The structures are fabricated using a commercial PCB fabrication process, and far-field angle-resolved measurements are used to verify the improved bandwidth of our design relative to a reference narrowband structure.


\section{Design of Broadband Metasurface for Anomalous Reflection}

\subsection{Achromatic and Dispersive Metasurface Designs \label{section:surface_imp}}

Here we consider two general approaches to generating broadband anomalous reflection using ultrathin metasurfaces. The first approach, the broadband achromatic design, is depicted in  Fig.~\ref{fig:ms_illustration}(a). The incident beam impinging upon the metasurface at an angle of $\theta_\mathrm{i}$ and reflected at an angle of $\theta_\mathrm{o}$  requires the metasurface to provide a phase discontinuity $\Phi_{r}$ which linearly changes with position $x$ and  frequency $\omega$, formulated as \cite{yu2011light},
\begin{equation}
\Phi_{r}(x,\omega)=\frac{\omega x \Delta_{\theta}}{c} + \Phi_0(\omega).    
\label{eq:phi_r}
\end{equation}
Here, $c$ is the speed of light in vacuum, $\Delta_\theta=\sin \theta_\mathrm{o}-\sin \theta_\mathrm{i}$, and $\Phi_0(\omega)$ is an arbitrary additional phase term independent of $x$, which does not affect the reflection angle. To assist with further calculation, the additional phase term is parameterized as $\Phi_0(\omega)=\omega t_0$, where $t_0$ represents an additional group delay.  
In the design of a broadband achromatic metasurface, in addition to a center frequency $\omega_0$, a certain fractional bandwidth $\Delta \omega$ is chosen. As illustrated by Fig.~\ref{fig:ms_illustration}(a), a broadband incident beam is reflected into the same angle at all frequencies. This operation requires the metasurface to have a distinct phase profile for every frequency within its bandwidth of interest, as shown in Fig.~\ref{fig:ms_illustration}(c). Therefore, broadband achromatic metasurfaces require a non-periodic structure and become increasingly difficult to realize for large bandwidths and aperture sizes. 


Alternatively, using the second approach, broadband metasurfaces can be designed to direct energy into a chosen diffraction order with high efficiency over some bandwidth, corresponding to a frequency-dependent reflection angle, as shown in Fig.~\ref{fig:ms_illustration}(b).
In this approach, the metasurface is designed to satisfy Eq.~\eqref{eq:phi_r} at a chosen center frequency $\omega_0=2\pi f_0$, and to maintain the same phase profile over the desired bandwidth $\Delta \omega$. This leads to a periodic reflection phase, with period $X$ given by
\begin{equation}
X=\frac{\lambda_0}{\sin \theta_\mathrm{o}-\sin \theta_\mathrm{i}},
\end{equation}
where $\lambda_0$ is the wavelength at the center frequency. 
The reflection phase of this dispersive design can then be expressed as
%
\begin{equation}
\begin{aligned}
\Phi_{r}(x,\omega)&=2\pi \frac{x}{X}+ \Phi_0(\omega)        
\end{aligned}
\label{eq:phi_r_periodic}
\end{equation}
%
This local reflection phase profile is shown in Fig.~\ref{fig:ms_illustration}(d). We see that the required phase gradient is no longer dependent upon frequency, and the group delay imparted by the phase profile is constant across all positions. Throughout this paper, this second approach is referred to as broadband dispersive design. 


To implement the phase profiles given in Eqs.~\eqref{eq:phi_r} and \eqref{eq:phi_r_periodic} we consider a meta-atom consisting of a single patterned metallic layer, separated from a metallic ground plane by a dielectric substrate. The interaction of the incident and reflected waves with the metasurface can be represented by a surface impedance distribution, accounting for the ratio of electric and magnetic field components tangential to the metasurface. We consider the case of TE polarized incident and reflected waves, with electric field in the $y$ direction. The relationship between the phase profile and the surface impedance can be specified as \cite{asadchy2016perfect}:
\begin{equation}
\begin{aligned}
\mathrm{Z_{surf}}(x,\omega)=-j{Z_o}\cot(\Phi_r/2), \\
\end{aligned}
\label{eq:surf_imp}
\end{equation}
where $Z_o=\eta /\cos\theta_\mathrm{o}$ is the impedance of the reflected wave and $\eta$ refers to the free space impedance. An equivalent circuit can be introduced to obtain the required impedance of the patterned metallic layer, facilitating the unit-cell design process. As shown in Fig.~\ref{fig:ms_illustration}(e), each meta-atom is described as a shunt impedance $Z_\mathrm{ms}$ loading a transmission line, which represents propagation through a dielectric substrate of thickness $d$. The other port of the transmission line is grounded \cite{sima2018combining}, representing the continuous metallic ground plane. From the transmission line's ABCD transfer parameters, we can relate Eq.~\eqref{eq:surf_imp} to the impedance requirement of the patterned metallic layer, formulated as
%
%
%
%
\begin{equation}
Z_{\mathrm{ms}}=\frac{Z_o Z_s\tan(\omega t_s)\cot(\Phi_{r}/2)}{jZ_s\tan(\omega t_s) + jZ_o\cot(\Phi_{r}/2)}.
\label{eq:zms_t0_ts}
\end{equation}

\begin{figure}[t]
\centering
\subfloat{
\includegraphics[height=4cm]{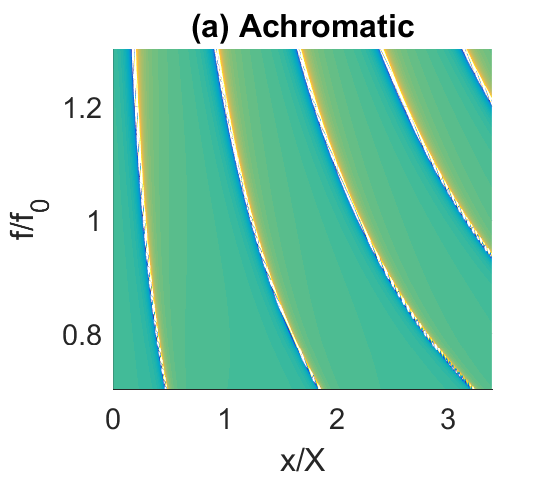}}
\subfloat{
\includegraphics[height=4cm]{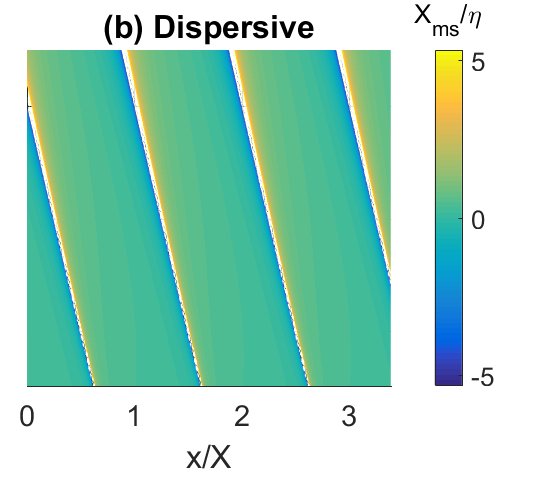}}
\hfil
\subfloat{
\includegraphics[width=\linewidth]{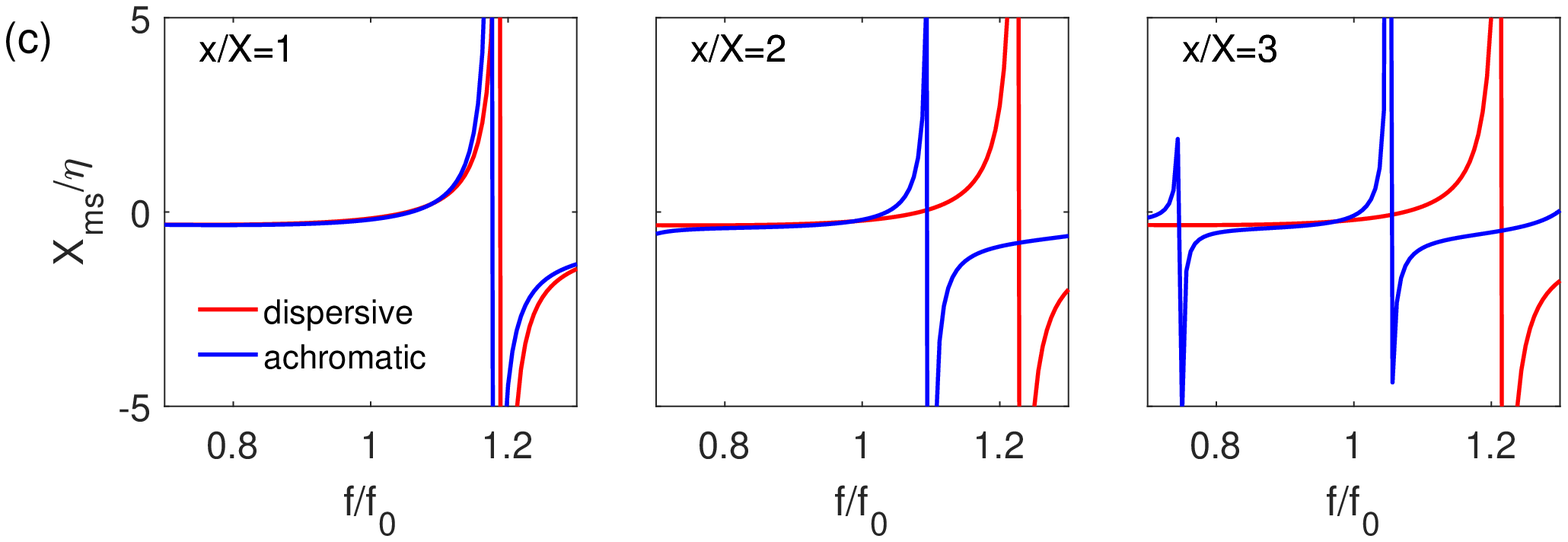}}
\caption{\label{fig:zms_x} The required meta-atom layer impedance as a function of both frequency $f$ and position $x$ for (a) broadband achromatic, and (b) broadband dispersive designs. (c) The required meta-atom layer impedance as a function of frequency for selected positions $x$ for broadband achromatic and broadband dispersive designs. $X=7.8\,mm$ is the period of metasurface for the center frequency $f_0=80\,\mathrm{GHz}$, with the designed anomalous reflection angle  $\theta_\mathrm{o}=30^\circ$ and normally incident illumination. The substrate is 0.254\,mm thick with a relative permittivity of 3.}
\end{figure}

Here, propagation through the substrate is expressed as a time delay $t_s=\sqrt{\epsilon_s} d/c$ and the substrate impedance is $Z_\mathrm{s}=\sqrt{\frac{\mu_0}{\epsilon_0\epsilon_s}}=\frac{\eta}{\sqrt{\epsilon_s}}$, and $\epsilon_0$ and $\mu_0$ are free space permittivity and permeability. We assume that the substrate thickness $d$ and permittivity $\epsilon_s$ are specified, based on the properties of available materials. 
The impedance given by Eq.~\eqref{eq:zms_t0_ts} is purely imaginary, and can be calculated over the entire  bandwidth of interest $\Delta\omega$. In Fig.~\ref{fig:zms_x}(a) and (b), we plot the reactance $X_{ms}=\mathrm{imag}(Z_{ms})$ as a function of both frequency and position, whereas in Fig.~\ref{fig:zms_x}(c) and (d), we plot $X_{ms}$ as a function of frequency at selected positions ($x/\mathrm{X}=1, 2, 3$).

Based on the stipulation of meta-atom layer impedances $Z_\mathrm{ms}$, a clear physical intuition of the required meta-atom geometry is be obtained. From $Z_\mathrm{ms}$ plotted in Fig.~\ref{fig:zms_x}(c)-(d), we see that resonant structures exhibiting poles and/or zeros in their impedance are required. In the achromatic case, the required number of resonators increases for larger values of $x$ (i.e.~for a larger aperture), as indicated by the increasing number of poles and zeros. In contrast, for the dispersive design, the number of poles and zeros in the operating band remains constant for any aperture size. 


\begin{figure*}
    \includegraphics[width=0.29\textwidth]{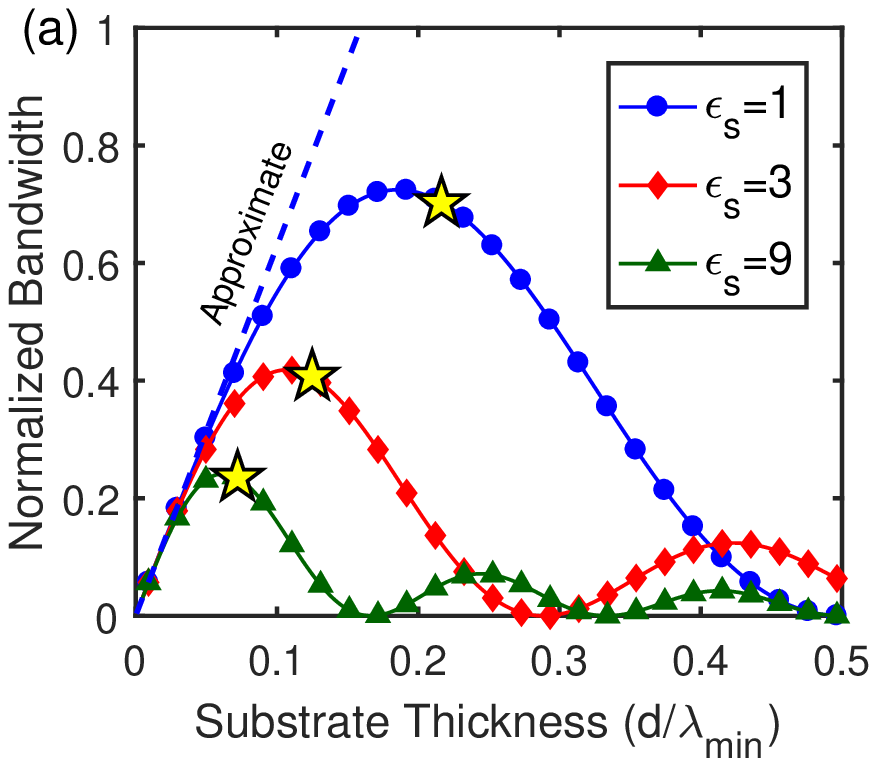} \quad
    \includegraphics[width=0.34\textwidth]{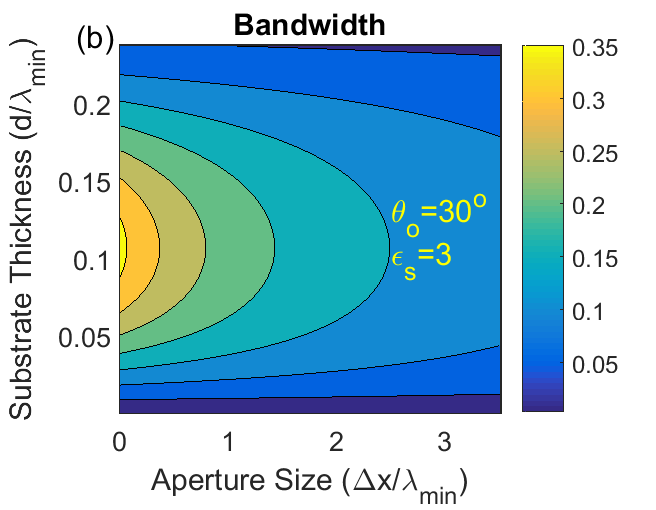} \quad
    \includegraphics[width=0.29\textwidth]{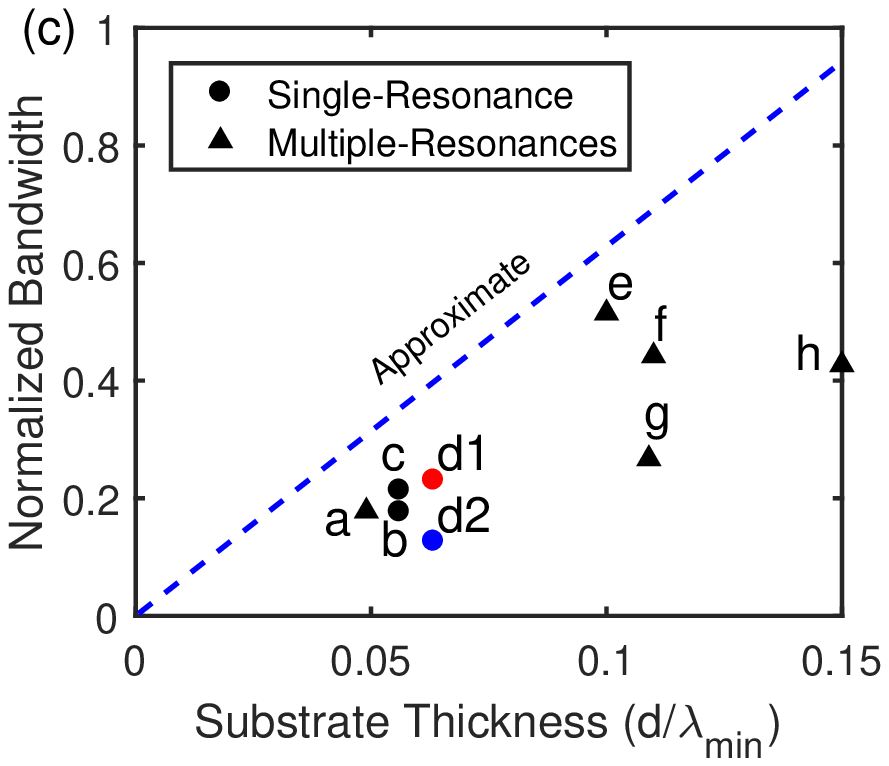}
\caption{\label{fig:bwlimit} (a) The bandwidth limit for a dispersive metasurface case, shown for three dielectric constants ($\epsilon_s$). It is normalized to the output angle as $\Delta \omega/(\omega_{min}\cos\theta_o)$, and plotted as a function of substrate thickness $d/\lambda_{min}$. The dashed line gives the approximate bandwidth limit from Eq.~\eqref{eq:bw_limit_d_simp} and the star markers indicate the the approximate thickness for maximum bandwidth $d_{peak}$ from Eq.~\eqref{eq:d_peak}. (b) The bandwidth limit in the achromatic metasurface case for a fixed substrate permittivity $\epsilon_s=3$ and output angle $\theta_o=30^\circ$. (c) A comparison of dispersive broadband structures reported in the literatures (see Table ~\ref{tab:ms_compare_bwlimit}) with the approximate limit from Eq.~\eqref{eq:bw_limit_d_simp}. Circles indicate meta-atoms with single resonances and triangles indicate meta-atoms with multiple-resonances. The red circle (d1) is the experimental result for our broadband design and the blue circle (d2) is for our reference narrowband design, as outlined in Section \ref{section:experiment}.}
\end{figure*}

\subsection{Bandwidth Limits of Single Resonance Meta-Atoms} \label{sec:delay_realizability}


The use of a single resonance meta-atom limits the bandwidth achievable, since the constraints of passivity and causality prevent it to exhibit arbitrary variation of impedance $Z_{ms}$ with frequency. In Ref.~\onlinecite{fathnan2018bandwidth} we previously derived bandwidth limits for achromatic metasurfaces based on such arguments, however the influence of substrate properties was neglected.
%
Following the approach of Ref.~\onlinecite{presutti2020focusing}, we make use of the time-bandwidth relationship for a single Lorentzian resonance. When combined with the physical constraints of passivity and causality, the time-bandwidth relationship can be used in defining more specific bandwidth limits, including quantifying the influence of substrate parameters. 

Derived from the single resonance Lorentzian model and the temporal coupling mode theory, the coupling lifetime of the resonance $\Delta T$ and the stored energy bandwidth $\Delta \omega$ are related by \cite{mann2019nonreciprocal}
\begin{equation}
    \Delta \omega \Delta T=2. \label{eq:tbp}
\end{equation}
In a meta-atom based on surface impedance approach, this time-bandwidth restriction describes the limited ability of a structure to match the required broadband surface impedance. This limitation is inherent within a single resonance structure since it can only exhibit a single pole or zero of impedance within the considered bandwidth. 
From Eq.~\eqref{eq:tbp}, the bandwidth limit can be obtained when  resonance lifetime $\Delta T$ is specified. 
As detailed in Section I of the Supplementary Material, we confirm that the resonance lifetime is equivalent to the phase slope or group delay imparted by the meta-atoms, written as 
\begin{equation}
    \Delta T= \bigg|\frac{\mathrm{d}\Phi_r}{\mathrm{d}\omega}\bigg|,
\end{equation}
Therefore, from the time-bandwidth product we obtain bandwidth limits for dispersive and achromatic metasurfaces as
\begin{align}
\Delta \omega &\leq \frac{2}{\,t_0} \qquad &\textrm{(Dispersive),}\label{eq:bw_resonator_d} \\
\Delta \omega &\leq \frac{2}{\,(t_0+\Delta_\mathrm{\theta}\Delta x/c)} \qquad &\textrm{(Achromatic).} \label{eq:bw_resonator_a}   
\end{align}
We see that the limits for single resonance metasurfaces are simple relations between the bandwidth and the additional group delay parameter $t_0$ that appears in the design. In the achromatic metasurface case, in addition to $t_0$, the group delay is also dependent on the aperture size ($\Delta x=x_\mathrm{max}-x_\mathrm{min}$), hence the limit is more stringent.

Since the additional group delay $t_0$ is a limiting factor on achievable bandwidth for both achromatic and dispersive metasurfaces, we must find the physical constraints on this value. From the meta-atom impedances requirement in Eq.~\eqref{eq:zms_t0_ts}, we can derive the minimum additional group delay by considering that they should comply with Foster's reactance theorem, i.e.~$\frac{\mathrm{d }}{\mathrm{d}\omega}X_{\mathrm{ms}}\geq 0$. This is equivalent to defining a minimum additional phase $\Phi_0$ for a passive, causal and loss-less realization, which was recognized in previous works on broadband metasurfaces \cite{aieta2015multiwavelength,cheng2019broadband,qu2015controlling}. 
%
%
%
As detailed in Section II of the Supplementary Materials, the minimum additional group delay $t_0$ for both the achromatic and the dispersive cases is
\begin{equation}
\label{eq:fm}
t_0=\frac{2Z_o t_s}{Z_s \sin^2(\omega_{min} t_s)},
\end{equation}
%
where $\omega_{min}$ is the minimum operating frequency for a broadband metasurface with $\Delta \omega=\omega_{max}-\omega_{min}$. Substituting the above equation into the achievable bandwidth in Eqs.~\eqref{eq:bw_resonator_d} and \eqref{eq:bw_resonator_a}, we obtain the limits on the bandwidth relative to  the minimum operating frequency. 
%
\begin{align}
\frac{\Delta \omega}{\omega_\mathrm{min}}&\leq \frac{Z_s \sin^2(\omega_{min} t_s)}{Z_o \omega_\mathrm{min} t_s} \qquad &\textrm{(Dispersive),} \label{eq:bw_limit_d}\\
\frac{\Delta \omega}{\omega_\mathrm{min}}&\leq \frac{2}{\frac{Z_o \omega_\mathrm{min} t_s}{Z_s \sin^2(\omega_{min} t_s)}+2\pi\frac{\Delta_x}{\lambda_\mathrm{min}}  \Delta_\theta }  &\textrm{(Achromatic)}. \label{eq:bw_limit_a}
\end{align}
%
%

In Fig.~\ref{fig:bwlimit}(a) we plot the bandwidth limit given by Eq.~\eqref{eq:bw_limit_d} for the dispersive case as a function of normalized substrate thickness $d/\lambda_\mathrm{min}$, for several values of substrate permittivity $\epsilon_s$. Here, $\lambda_\mathrm{min}$ refers to the wavelength at the minimum operating frequency ($2\pi c/\omega_\mathrm{min}$). To show results for all reflection angles, we incorporate the output angle $\theta_o$ into a normalized bandwidth $\Delta\omega/(\omega_\mathrm{min}\cos\theta_o)$. In Fig.~\ref{fig:bwlimit}(b), the bandwidth limit $\Delta\omega/\omega_\mathrm{min}$ is shown for the achromatic case. In contrast to the dispersive case, the aperture size $\Delta x$ plays a role, so we plot the bandwidth as a function of both aperture size $\Delta x/\lambda_\mathrm{min}$ and substrate thickness  $d/\lambda_\mathrm{min}$. Note that the bandwidth limit in Fig.~\ref{fig:bwlimit}(b) is shown for a fixed value of output angle and substrate permittivity.

We see that both Eq.~\eqref{eq:bw_limit_d} and \eqref{eq:bw_limit_a} give limits inversely proportional to the substrate thickness, therefore a broadband metasurface cannot be infinitesimally thin. However, both equations include a sine-squared function of substrate thickness, meaning that substrate thickness cannot be increased arbitrarily to increase bandwidth. From Fig.~\ref{fig:bwlimit}(a) we see that the bandwidth oscillates with the thickness, and has a maximum value that depends on the substrate permittivity. The maximum value can also be seen for achromatic metasurfaces as shown by Fig.~\ref{fig:bwlimit}(b). We have confirmed that the peak for the achromatic case occurs at the same thickness value as in the dispersive case. The maximum bandwidth is obtained when the thickness reaches $d_{peak}$, corresponding to the first peak of the oscillation. Since it is clear that the peak comes from the sine function, we expect the peak of the full limit to occur when $\omega_{min} t_s \approx \frac{\pi}{2}$, leading to
\begin{equation}
    d_{peak}\approx\frac{\lambda_{min}}{4\sqrt{\epsilon_s}} \label{eq:d_peak}
\end{equation}
Equation \eqref{eq:d_peak} predicts that the maximum bandwidth is obtained when the substrate's thickness is a quarter wavelength in the substrate material. For each curve in  Fig.~\ref{fig:bwlimit}(a), $d_\mathrm{peak}$ is plotted with a star marker, corresponding closely to the peak. This result is consistent with previous studies on similar planar metallic architectures, showing that magnetic coupling begins to reduce beyond a quarter wavelength thickness \cite{pu2013broadband}.

\begin{table*}[t]
\caption{\label{tab:ms_compare_bwlimit} Comparison of various types of resonance implementation for broadband dispersive metasurfaces.}
\begin{ruledtabular}
\begin{tabular}{lcccccccccc}
 Column (1)& (2) & (3) & (4) & (5) & (6) & (7) & (8) & (9) & (10) & (11)\\
               Meta-atom structures & Resonance type & $\theta_o$ & $\lambda_{min}$ &  $d$         & $\epsilon_s$ &$d/\lambda_{min}$ & $d_{peak}/\lambda_{min}$ & $\frac{\Delta \omega}{\omega_{min}}$ & $\frac{\Delta \omega/\omega_{min}}{\cos(\theta_o)}$ &  $\frac{\Delta \omega/\omega_{min}}{\cos(\theta_o)(d/\lambda_{min})}$ \\
\hline
{a.   Single dipoles \& loops \cite{chen2013x}} &Multiple  & 30                    & 30.769   mm     & 1.500   mm          & 2.20 & 0.049    & 0.17       & 0.154                              & 0.178                       & 3.65                                      \\
{b.   Nano-bricks \cite{sun2012high}}   & Single            & 45           & 900   nm        & 50   nm             &4.87 & 0.056     & 0.11           & 0.125                              & 0.177                       & 3.18                                      \\
{c.   Nano-bricks \& crosses \cite{pors2013plasmonic}}    & Single & 33.7       & 900   nm        & 50   nm           & 3.90   & 0.056    & 0.13            & 0.178                              & 0.214                       & 3.85                                      \\
{d1.   LC resonances (\textbf{this work})}      & Single      & 30               & 4   mm          & 0.   254 mm     & 3.00     & 0.064 & 0.14               & 0.200                              & 0.231                       & 3.64                                      \\
{e.   Multiple loops \& dipoles \cite{qu2015controlling}}    & Multiple      & 15          & 1.5   mm        & 0.120   mm      & 2.50     & 0.080      &0.16          & 0.500                                & 0.518                       & 6.40                                       \\
{f.   Stub-loaded dipoles \cite{YouTerahertzreflectarrayenhanced2019}} & Multiple      & 9.5        & 0.33   mm       & 36.000   $\mu$m   & 2.33       & 0.109         & 0.16       & 0.264                              & 0.267                       & 2.45                                      \\
{g.   Stacked dipoles \& rings \cite{qu2016wideband}} & Multiple & 34.61     & 27.27   mm      & 3.500   mm & 2.20           & 0.128      & 0.17          & 0.364                              & 0.442                       & 3.44                                      \\
{h.   Five parallel dipoles \cite{yoon2015broadband}}    & Multiple & 19.4        & 21.23   mm      & 3.175   mm       & 2.20    & 0.150  & 0.17              & 0.403                              & 0.427        & 2.85    \\ 
\end{tabular}
\end{ruledtabular}
\end{table*}

Since the maximum bandwidth in Eqs.~\eqref{eq:bw_limit_d} and \eqref{eq:bw_limit_a} occurs at low to moderate values of substrate thickness and permittivity, we can simplify these expressions by considering $d$ and $\epsilon_s$ to be small (i.e.~$\omega_\mathrm{min} t_s \ll 1$). Approximating the sine function by its first order Taylor series yields
\begin{align}
\frac{\Delta \omega}{\omega_\mathrm{min}}&\leq 2\pi\frac{d}{\lambda_\mathrm{min}}\cos\theta_{o} \qquad &\textrm{(Dispersive),} \label{eq:bw_limit_d_simp}\\
\frac{\Delta \omega}{\omega_\mathrm{min}}&\leq \frac{2}{\frac{\lambda_\mathrm{min}}{d}\frac{1}{\pi\cos\theta_{o}}+2\pi\frac{\Delta_x}{\lambda_\mathrm{min}}  \Delta_\theta }  &\textrm{(Achromatic)}. \label{eq:bw_limit_a_simp}
\end{align}

Equations \eqref{eq:bw_limit_d_simp} and \eqref{eq:bw_limit_a_simp} give simple relations between bandwidth, thickness, reflection angle, and (for the achromatic case) aperture size. Note that the dispersive limit in Eq.~\eqref{eq:bw_limit_d_simp} has a similar form to the limits previously derived for other broadband composite planar structures \cite{gustafsson2011physical,brewitt2007limitation}. However, Eq.~\eqref{eq:bw_limit_d_simp} also includes the influence of the reflection angle which is specific for metasurfaces involving wavefront manipulations. The approximate limit for dispersive metasurfaces is plotted as a dashed line in Fig.~\ref{fig:bwlimit}(a), showing good agreement with the exact expressions. Consistent with previous studies, we note that this approximate expression is independent of the substrate permittivity $\epsilon_s$. However, the validity of the approximate expression becomes worse for higher value of substrate permittivity. 
Although a thicker and lower substrate permittivity has the highest possible bandwidth, it should also be noted that a very low dielectric constant makes it more difficult to realize meta-atoms with highly sub-wavelength resonances.

In Fig.~\ref{fig:bwlimit}(c) we compare the performance of several dispersive metasurfaces reported in the literature \cite{chen2013x,sun2012high,pors2013plasmonic,qu2015controlling,YouTerahertzreflectarrayenhanced2019,qu2016wideband,yoon2015broadband} with our approximate limit (dashed line). To obtain fair comparison over different reflection angles, we plot the normalized bandwidth $\Delta\omega/(\omega_\mathrm{min}\cos\theta_o)$. Details of each structure are outlined in Table \ref{tab:ms_compare_bwlimit}. This comparison shows that larger thickness can allow a more broadband metasurface, while as expected, none of the presented metasurfaces use thickness larger than the quarter wavelength limit (see Column 8 of Table \ref{tab:ms_compare_bwlimit}). We also notice that even though some of the presented metasurfaces are based on multi-resonance structures, when we normalized the contribution of thickness, in some cases their bandwidth performance is exceeded by single resonance structures (see Column 11 of Table \ref{tab:ms_compare_bwlimit}).
In Section \ref{section:experiment} we design a metasurface based on appropriate tailoring of single resonance structures that can yield wide bandwidth without requiring multi-resonance structures. The measured bandwidth of our broadband metasurface is plotted as a red dot (d1) in Fig.~\ref{fig:bwlimit}(a). We see that the bandwidth of d1 approaches g and exceeds a, both of which use more complicated multiresonance structures for their realizations. For comparison purposes we also show results for our reference narrowband structure with the blue dot (d2).




\subsection{\label{sec_realization_bb_imp} Realization of Broadband Impedance}

Having determined the meta-atom layer impedance profile for broadband operation, it is required to translate the abstract impedance functions into realistic metallic structures. Several works have shown this impedance translation procedure, however, they rely on numerical optimization tools and consider only narrowband operation \cite{epstein2016huygens,pfeiffer2013millimeter}. To efficiently realize a broadband metasurface, a more general approach for fitting the impedance over a broad frequency range is required.
Here, we introduce a method of fitting the surface impedance to an LC resonant circuit. By evaluating the required surface impedance, we can decide a suitable implementation of the meta-atom LC resonance based on how close it is to the pole or zero of impedance $\mathrm{Z_{ms}}$. Further details on this point are given in Section III of the Supplementary Material. 


To facilitate the design of realistic structures, we find the series inductance ($L_s$) and capacitance ($C_s$) required for the LC-series implementation. The general solution for the LC-series reactance is
\begin{equation}
\begin{aligned}
    \mathrm{X({\omega_0})}=\omega_0 L_s-\dfrac{1}{\omega_0 C_s} \\
\end{aligned}
\label{eq:xms}
\end{equation}
$\mathrm{X({\omega_0})}$ is the imaginary part of impedance at the center frequency ($\omega_0$) stipulated by Eq.~\eqref{eq:zms_t0_ts}. To obtain a solution for two variables ($L_s$ and $C_s$) an additional equation is needed. From Eq.~\ref{eq:xms} we can calculate the derivative of the reactance at $\omega_0$, which is related to the inductance and capacitance as
\begin{equation}
\begin{aligned}
    \frac{\mathrm{d}X({\omega_0})}{\mathrm{d}\omega_0}=\dfrac{1}{\omega_0^2C_s}+L_s
\end{aligned}
\label{eq:xms_deriv}
\end{equation}
Solving equations \eqref{eq:xms} and \eqref{eq:xms_deriv} gives the following solutions for the series LC circuit elements 
\begin{align}
L_s &=\frac{X(\omega_0) + \omega_0 X'(\omega_0)}{2\omega_0} \label{eq:inductance_derv} \\
C_s &=\frac{-2}{\omega_0(X(\omega_0) - \omega_0 X'(\omega_0))} \label{eq:capacitance_derv}
\end{align}
For impedance functions that are better fitted by a parallel LC resonance, expressions for the inductance $L_p$ and capacitance $C_p$ are obtained by stipulating susceptances ($B_\mathrm{ms}=-\frac{1}{X_\mathrm{ms}}$), where the poles of impedance are transformed into zeros. The susceptance is fitted to a parallel LC resonance in a similar manner to the series case, yielding 
\begin{align}
C_p &=\frac{B(\omega_0) + \omega_0 B'(\omega_0)}{2\omega_0} \\
L_p &=\frac{-2}{\omega_0(B(\omega_0) - \omega_0 B'(\omega_0))}
\end{align}

\begin{figure}[t]
	\includegraphics[width=\linewidth]{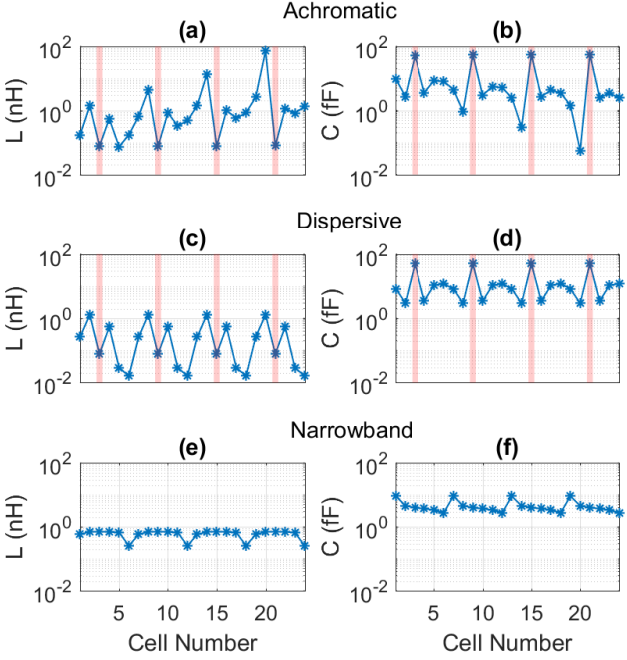}
	\caption{\label{fig:LC_all} Required inductance and capacitance values (on a log-scale) for different locations across each metasurface. (a,b) Broadband achromatic design, (c,d) broadband dispersive design, (e,f) narrowband design. The red shading indicates a cell with a parallel LC circuit implementation, the remaining cells are implemented using a series LC circuit.}
\end{figure}

To design a benchmark narrowband structure, we work directly with a realistic structure to match the phase requirement at a single frequency. This is equivalent to changing L and C values to match the impedance in Eq.~\eqref{eq:xms} but not its derivative in Eq.~\eqref{eq:xms_deriv}. 
In Fig.~\ref{fig:LC_all}, the fitted inductance and capacitance for the broadband achromatic metasurface (a,b), are compared to the broadband dispersive (c,d) and the narrowband case (e,f). We use the design parameters outlined in Section \ref{section:surface_imp} with period of X=7.8\,mm and six meta-atoms per super-cell for the periodic designs (meta-atom lateral size=1.3\,mm). Meta-atoms which require LC-parallel circuits for their realization are indicated by red shading. 

In Fig.~\ref{fig:LC_all}, we see that for the achromatic metasurface, the required LC values are not periodic. For distances further from the center, the inductance increases while the capacitance decreases. For both inductance and capacitance, the ratio between the smallest and largest values exceeds $10^3$. This makes achromatic metasurfaces with a large aperture very difficult to realize, since fabrication tolerances and available space within each meta-atom's cell will limit the achievable values of inductance and capacitance. In the broadband dispersive metasurface, the required LC values are periodic, with maximum ratio between the smallest and the largest below $10^2$. This makes the broadband dispersive metasurface much easier to realize than the broadband achromatic one. In the narrowband case, since the derivative of impedance is not taken into account, only small changes in either inductance or capacitance are needed, as depicted in Fig.~\ref{fig:LC_all}(e,f). We will show in the next section that this can be realized by changing only one geometrical parameter of our chosen metallic structure. 




\section{Synthesis of Single Resonance Broadband Metasurfaces \label{section:experiment}}

\begin{figure}[b]
{\includegraphics[width=0.98\linewidth]{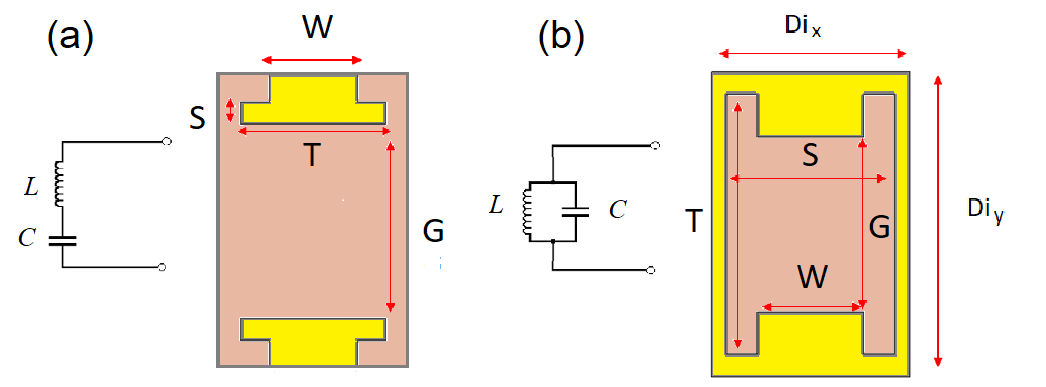}}
	\includegraphics[width=0.49\linewidth]{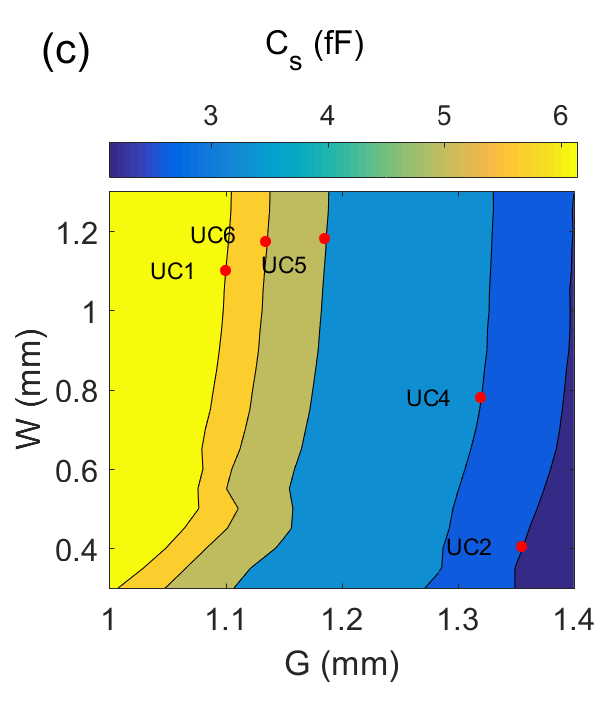}
    \includegraphics[width=0.49\linewidth]{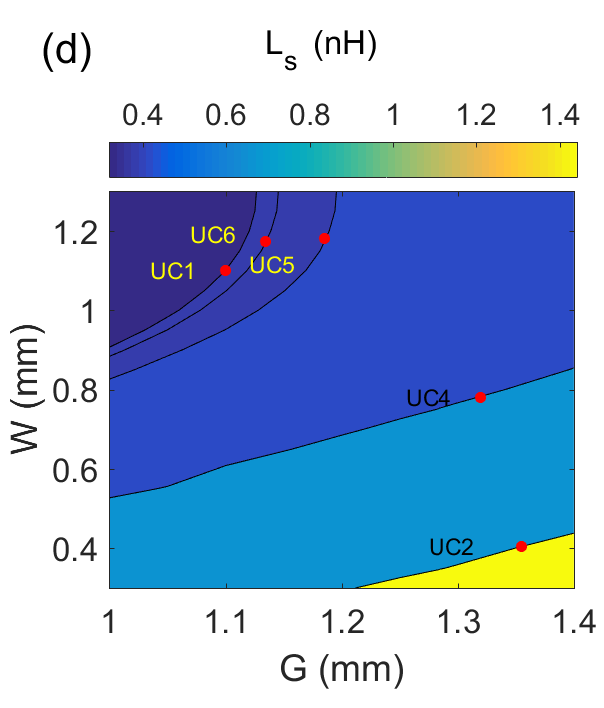}
\caption{\label{fig:uc_circ} Metasurface unit cell realization: (a) dog-bone structure, having a series LC resonance (b) inverse dog-bone structure, having a parallel LC resonance. (c) Capacitance and (d) inductance, extracted from the dogbone structure using numerical simulation with varying parameter G and W. Red dots indicate the chosen dog-bone implementation.} 
\end{figure}

\begin{figure*}[t!]
	\subfloat{\includegraphics[width=0.3\linewidth]{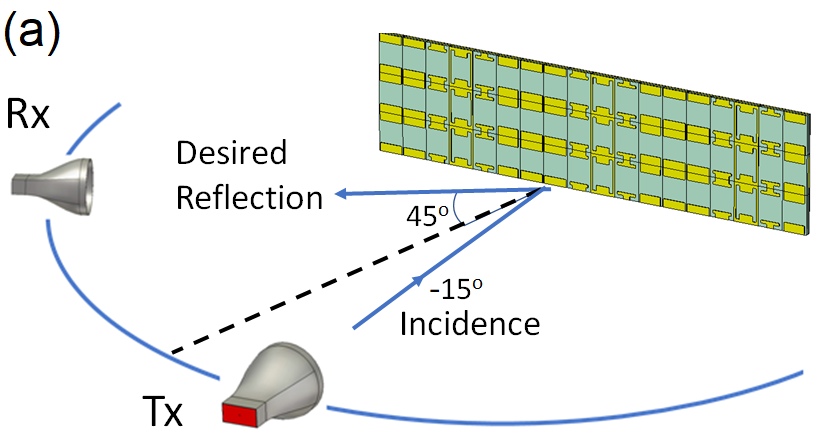}} \quad
	\subfloat{\includegraphics[width=0.33\linewidth]{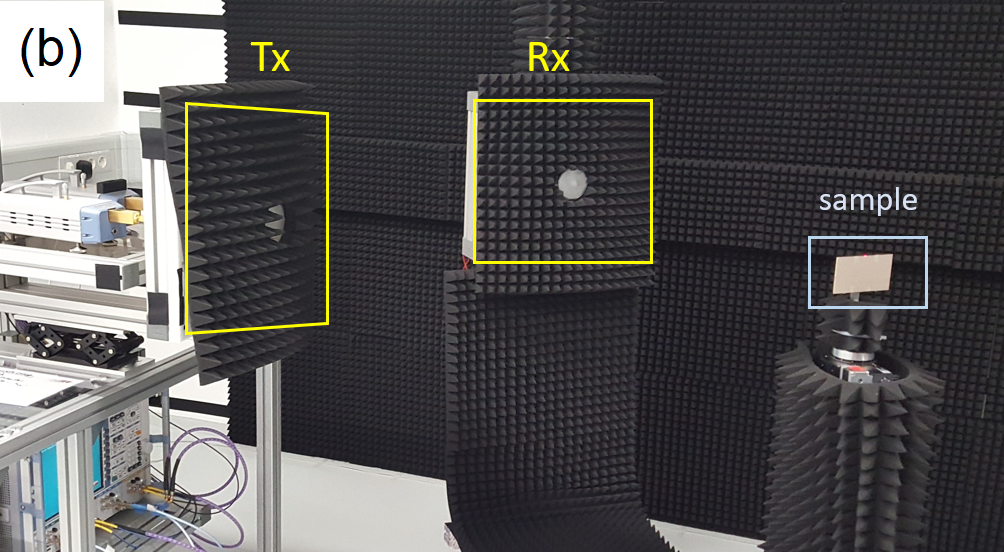}}
	\quad
	\subfloat{\includegraphics[width=0.3\linewidth]{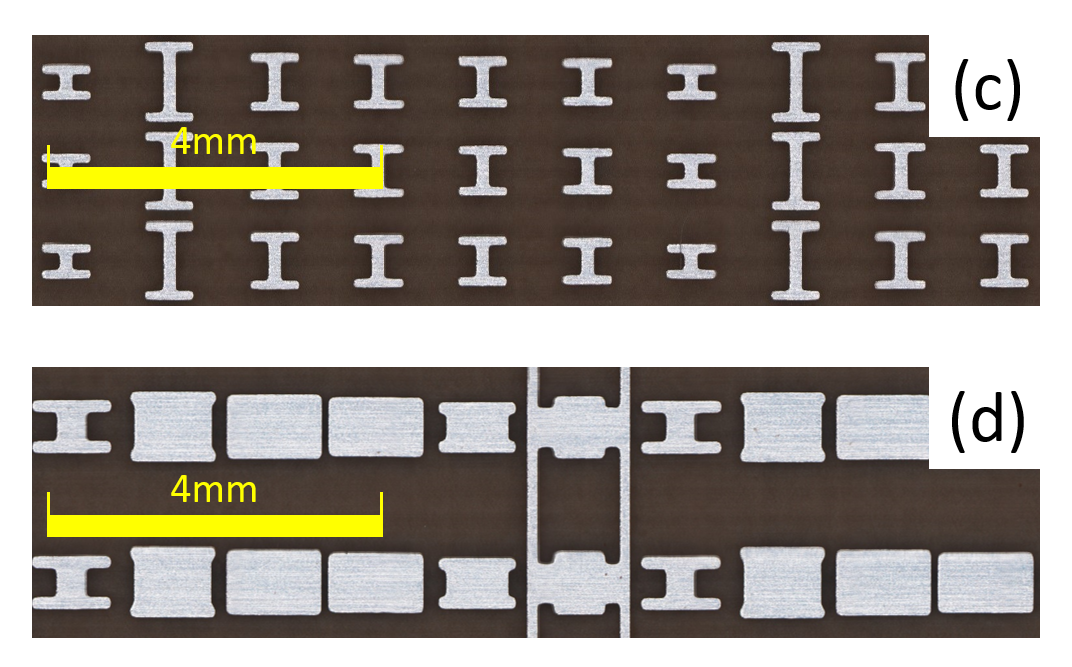}}
	\hfil
	\\
	\subfloat{\includegraphics[width=0.25\linewidth]{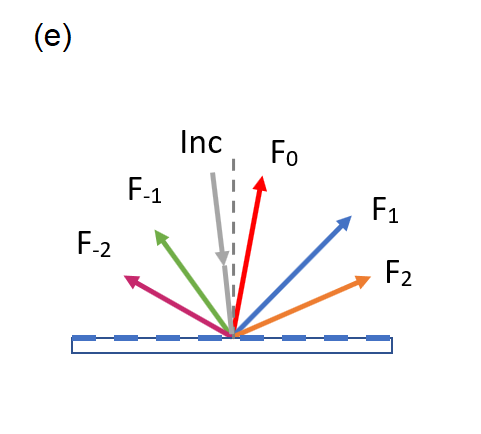}}
	\subfloat{\includegraphics[width=0.33\linewidth]{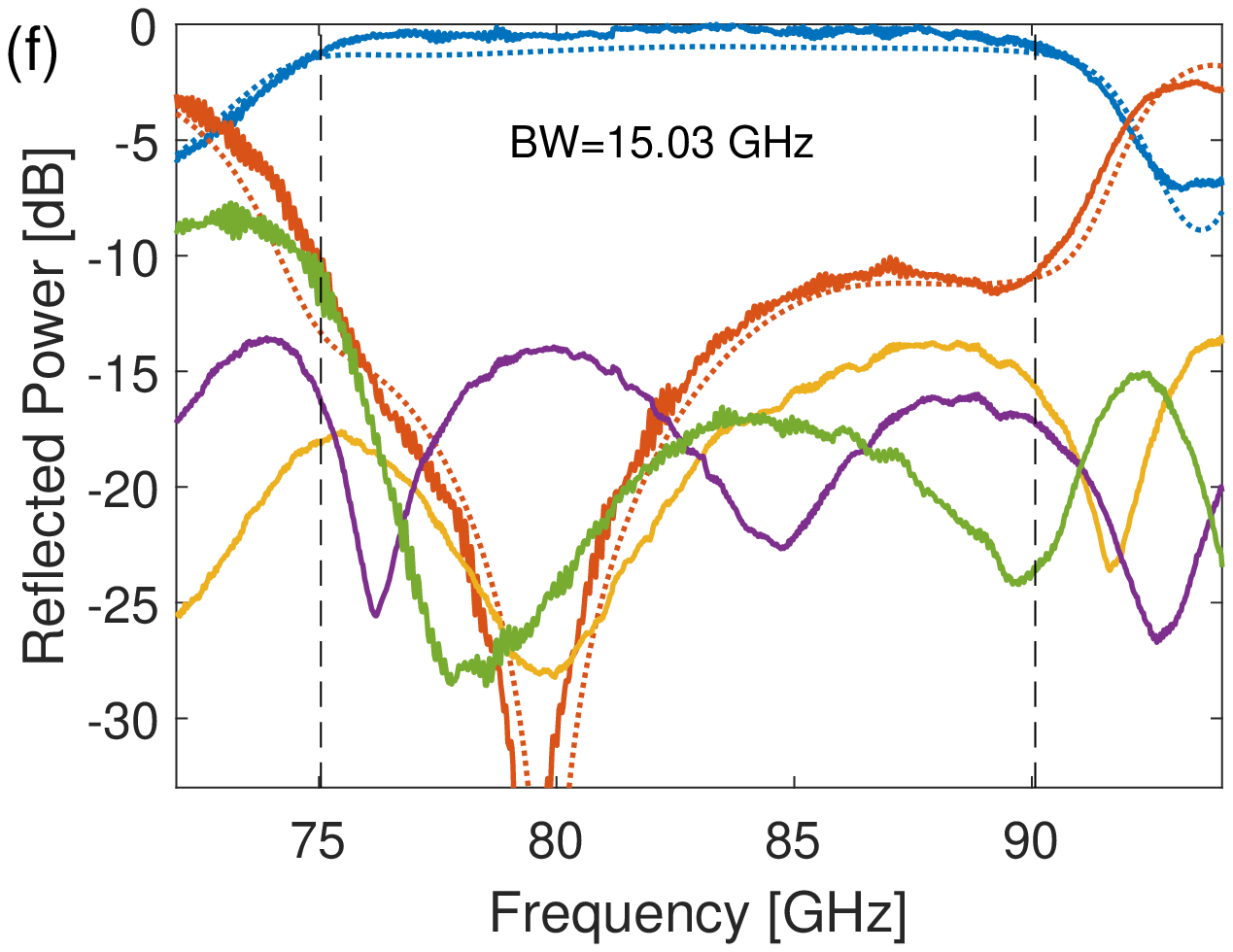}} \qquad
    \subfloat{\includegraphics[width=0.33\linewidth]{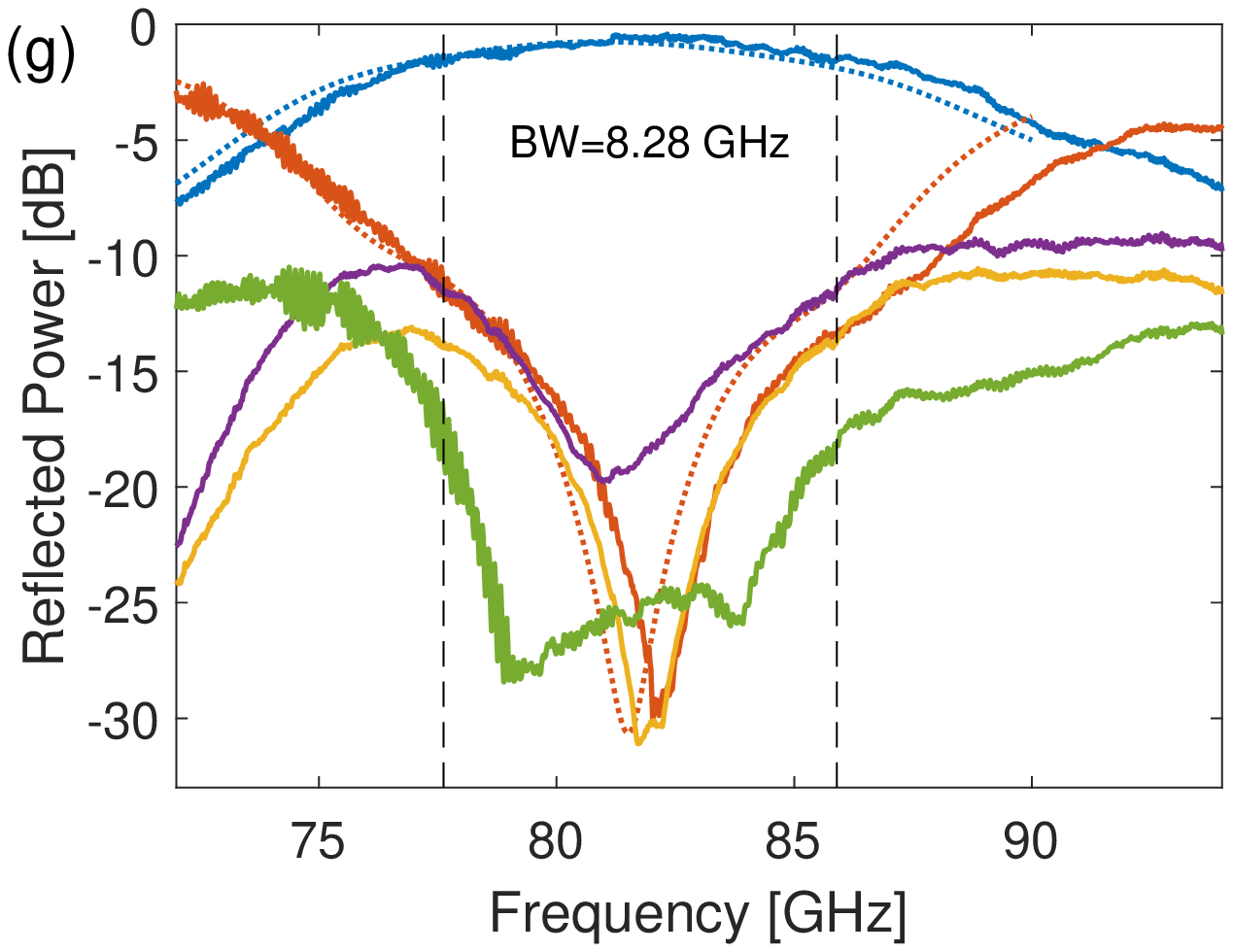}}
\caption{\label{fig:sample_meas}(a) Illustration of bi-static measurement with transmitter (Tx) and receiver (Rx) arms. (b) Photograph of the measurement setup. (c) Zoomed photograph of the narrowband metasurface sample and (d) the broadband dispersive sample. (e) Schematic of all propagating Floquet harmonics that occur in the measurement, with the results over the frequency range of 72-94\,GHz are plotted in (f) for the broadband dispersive design, and (g) for the narrowband design. Identical colors are indicated in (e) for each Floquet harmonics and simulations are plotted in  dotted-lines.}
\end{figure*}

\subsection{Meta Atom Design} \label{sec:metaatom_design}

To implement the series LC resonances we use dog-bone structures and for the parallel resonances we use inverse dog-bone structures. As shown in Fig.~\ref{fig:uc_circ}(a), the dog-bone structure is an anisotropic capacitively loaded dipole \cite{MunkFrequencySelectiveSurfaces2000,pfeiffer2013millimeter}, whereas the inverse dog-bone structure is an inductive grid with a small metallic inclusion \cite{holloway2018generalized} as shown in Fig.~\ref{fig:uc_circ}(b). 
Simulations are performed using the frequency domain solver of CST Microwave Studio. To calculate the response of each structure we use a locally homogeneous approximation, based on unit-cell boundary conditions. From the reflection parameters, we calculate the effective metasurface impedance $\mathrm{Z_\mathrm{surf}}$ using Eq.~\eqref{eq:surf_imp}. We then implement Eq.~\eqref{eq:zms_t0_ts} in a post-processing procedure to extract $\mathrm{Z_\mathrm{ms}}$ automatically from each simulation run. In this way, we obtain the impedance $\mathrm{Z_\mathrm{ms}}$ as well as its derivative at the operating frequency $f_0=80$\,GHz and can calculate inductance and capacitance according to Eqs.~\eqref{eq:inductance_derv} and \eqref{eq:capacitance_derv}. In this LC extraction, we use normal incidence ($\theta_{i} = 0^\circ$). We have confirmed that there is minimal deviation of L and C for $\theta_{i} \leq 30^\circ$, as detailed in Section IV of the Supplementary Material.

Fig.~\ref{fig:uc_circ}\,(c) and (d) depict how the dog-bone structure parameters $G$ and $W$ determine the capacitance $C_s$ and inductance $L_s$. The highlighted contours indicate capacitance and inductance suitable for a broadband dispersive design. The chosen $W$ and $G$ values for the corresponding unit cell implementations are indicated by red-dots and labeled by unit cell number $\mathrm{UCn}$. From both figures, we see that the minimum obtainable capacitance is slightly below 1\,fF, and the maximum obtainable inductance is around 1.4\,nH. Although these values are within the range for the broadband dispersive design, as shown in Fig.~\ref{fig:LC_all}\,(a,b), they do not cover the required inductance and capacitance of the achromatic design, even for a relatively small aperture of around 18 cells. Therefore, we only attempt to realize the broadband dispersive metasurface, as well as a narrowband metasurfaces to serve as a reference.

The narrowband design is implemented by changing the gap $G$, to modify capacitance and inductance in accordance with Fig.~\ref{fig:LC_all}\,(e, f). As can be seen from Fig.~\ref{fig:uc_circ}\,(c, d), varying $G$ drastically changes the capacitance but only slightly changes the inductance. On the other hand, varying $W$ has very small influence on the capacitance. Therefore, varying only the geometrical parameter $G$ can give full phase coverage at the center frequency, leading to the narrowband inductance and capacitance values plotted in Fig.~\ref{fig:LC_all}\,(e,f). The complete geometrical parameters for the broadband dispersive and narrowband metasurfaces are shown in the Section V of the Supplementary Material. 


\subsection{Numerical and Experimental Verification}

To verify the design procedure, we fabricate metasurface samples for operation within the W-band (75-110\,GHz) and measure their far-field responses. We use standard commercial printed circuit board (PCB) processing with an etching resolution of 100\,\textmu m. The substrate is Isola ASTRA MT77 which has a dielectric constant of 3, a dissipation factor of 0.0017 and a thickness of 254\,\textmu m. Both metallic layers are copper, with a thickness of 18\,\textmu m. The overall sample sizes are 100\,mm $\times$ 55\,mm. Both, the broadband dispersive and the narrowband designs were fabricated and a microscopy photograph of each design sample is shown in Fig.~\ref{fig:sample_meas}(c) and (d), respectively.   
A bi-static measurement setup \cite{olk2017highly} operating at W-band is used to characterize the metasurface samples, as shown in Fig.~\ref{fig:sample_meas}(a) and (b). The receiver arm (Rx) can move on a circle with a radius of 1 meter and the bistatic angle can be varied from  25$^\circ$ to 335$^\circ$. The intensity of the Floquet harmonics was determined using fine angular sweeps of the bistatic angle analogous to the procedure reported in \cite{olk2019experimental}. The procedure involves four runs of far-field measurement as detailed in Section VI of the Supplementary Material. The rotation has 0.1$^\circ$ precision, controlled automatically by an external computer which also collects measurement data from the Vector Network Analyzer (VNA). This angular precision is required as the direction of propagation of Floquet harmonics changes slightly within the frequency range of interest. 

\begin{figure}[t!]
\subfloat{\includegraphics[width=0.518\linewidth]{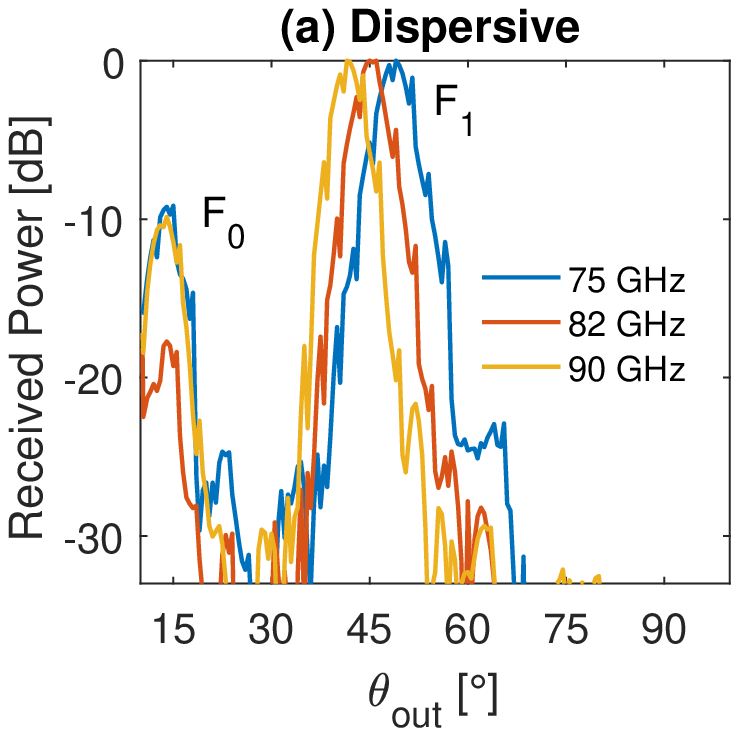}}
\hfil
\subfloat{\includegraphics[width=0.428\linewidth]{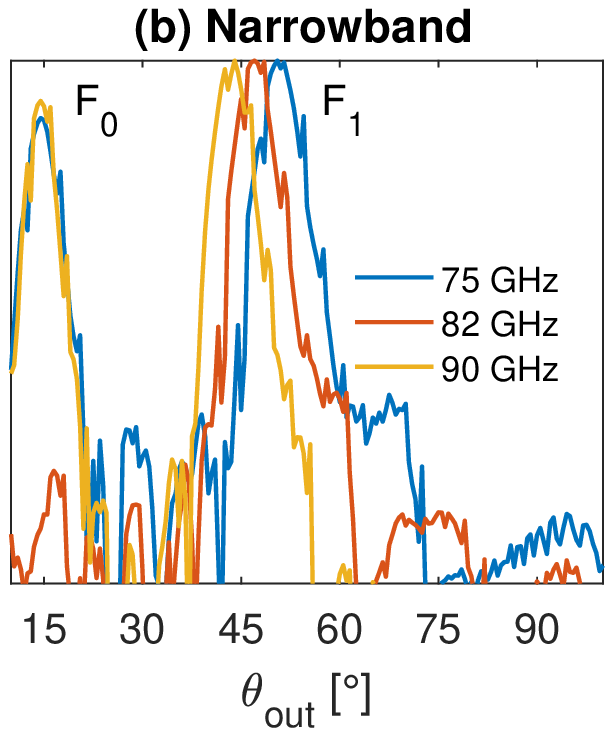}}
\caption{\label{fig:meas_res} Far field measurement results of $\mathrm{F_0}$ and $\mathrm{F_1}$ from (a) the broadband dispersive metasurface and (b) the narrowband metasurface for three different frequencies.}
\end{figure}

The propagating Floquet harmonics which may be reflected from the metasurface are illustrated in Fig.~\ref{fig:sample_meas}(e), and all of these were measured. The measurement results are plotted in Fig.~\ref{fig:sample_meas}(f) and (g) for the broadband dispersive and narrowband metasurfaces respectively. They are normalized to reflection spectra from a metallic mirror having the same size as the metasurfaces. We run a full-wave simulation using CST Microwave Studio and the results for the specular reflection ($\mathrm{F_0}$) and the desired diffraction order ($\mathrm{F_1}$) are plotted as dotted lines. The metasurfaces were designed for 80 GHz center frequency, however, we observe that the spectra are shifted by around 2 GHz due to fabrication tolerances. The simulation results plotted in Fig.~\ref{fig:sample_meas} were calculated with the patterned metallic layer scaled by 0.976 in the $x$ and $y$ axes, in order to have a good match with measurement results.

Measurements of both structures show that the dominant mode is the desired +1 diffraction order ($\mathrm{F_1}$). However, the broadband design has a more stable amplitude, maintaining a higher received power in a larger bandwidth. For the narrowband design, we observed a faster reduction in amplitude of the $\mathrm{F_1}$ mode as the frequency shifts further away from the center. Additionally, the narrowband design has a faster increase of undesired diffraction orders ($\mathrm{F_0}$ and $\mathrm{F_{-2}}$) away from the center frequency, which compromise the anomalous reflection performance. We calculate the bandwidth over which  the desired reflection spectra is above -3\,dB, with minimum 10\,dB difference to other diffraction modes. In the broadband design, the bandwidth is 15.03\,GHz corresponding to a fractional bandwidth ($\Delta \omega/\omega_{min}$) of 20.04\%, whereas in the narrowband design the bandwidth is 8.28\,GHz corresponding to a fractional bandwidth of 10.67\%. These results are included in Fig.~\ref{fig:bwlimit}(a) as a red dot (d1) for the broadband structure and blue dot (d2) for the narrowband structure. These results confirm that our structure is competitive with other designs, including some which make use of multiple resonances.


The measured far-field pattern of the broadband dispersive and narrowband metasurfaces are plotted in Fig.~\ref{fig:meas_res}(a)-(b) for three different frequencies. Each measurements is normalized to its peak value. The incident angle is -15$^\circ$ and the desired diffraction order is $\theta_\mathrm{o}=45^\circ$. 
The specular reflection peaks are at 15$^\circ$ and are indicated by $\mathrm{F_0}$. The broadband dispersive measurements show specular reflection equal to or below $-10$\,dB at all frequencies, whereas the narrowband metasurface has specular reflection of around $-3$\,dB at 75 and 90\,GHz. 

\section{Conclusion}
We have derived a limit on the achievable bandwidth for ultrathin reflective metasurfaces based on arguments that the meta-atoms exhibit single resonance and that they should be passive, causal and loss-less structures. Two scenarios of broadband metasurface were considered: the achromatic case, which maintains identical functionality with frequency, and the dispersive case, which gives a shift of beam angle, but is much simpler to realize. We reveal that in both cases there is a trade-off between substrate thickness and bandwidth in addition to the trade-off introduced by the angle of operation. The limit predicts a maximum achievable bandwidth for a substrate thickness of approximately a quarter wavelength, where beyond this thickness, the bandwidth degrades. We collected results from several other works in the literature, confirming that they fall within our limit. In the achromatic case, a more stringent limit applies, since the aperture size also contributes to the achievable bandwidth. 

Based on these fundamental relations, we developed a synthesis procedure for single resonance broadband metasurfaces. In realizing the metasurface, the required meta-atom layer impedance is fitted using either a series or parallel configuration of LC resonances, in which they are translated into realistic structures using dog-bone or inverse dog-bone structures. We verified the proposed method experimentally at millimeter-wave frequencies. The broadband dispersive metasurface achieves more than 90\% increase of bandwidth compared to the narrowband design, with 10\,dB contrast between the desired diffraction order and all other spurious orders. The bandwidth of our structure was shown to compare favourably with comparable broadband dispersive metasurfaces reported in the literature. Since the presented procedure to obtain broadband impedance is generic, it allows the implementation of other wave-front manipulation functions and is applicable to other planar metallic meta-atom geometries, including those suited for terahertz and infrared wavelengths.

\begin{acknowledgments}
This work was financially supported by the Australian Research Council (Linkage Project LP160100253), the  Luxembourg Ministry of the Economy (grant CVN 18/17/RED), the University of New South Wales (UIPA scholarship) and the Indonesia Endowment Fund for Education (LPDP) (PRJ-1081/LPDP.3/2017). The authors acknowledge fruitful discussions with Francesco Monticone.
\end{acknowledgments}

\bibliography{_main}

\setcounter{section}{0}
\setcounter{equation}{0}
\setcounter{figure}{0}
\setcounter{table}{0}
\renewcommand{\theequation}{S-\arabic{equation}}
\renewcommand{\thefigure}{S-\arabic{figure}}
\renewcommand{\thetable}{S-\Roman{table}}

\onecolumngrid
\clearpage
\section*{S\MakeLowercase{upplementary} M\MakeLowercase{aterial for}: \\ Bandwidth Limit and Synthesis Approach for Single Resonance Ultrathin Metasurfaces}


\section{\label{apx:res_lifetime} Resonance Lifetime and Meta-atom Group Delay}

In our single resonance meta-atom, the reflection spectrum can be described by a Lorentzian function \cite{mann2019nonreciprocal,siegman1986lasers}, and the time-evolution can also be described by temporal coupled mode theory \cite{suh2004temporal,haus1984waves}, formulated as
\begin{align}
|a|^2&=  \frac{1}{(\omega-\omega_0)^2-\gamma^2}, \label{eq:lorentzian} \\
a(t)&=a_0 \, \mathrm{exp}(i\omega_0t)\,\mathrm{exp}(-\gamma t). \label{eq:timeevolve} 
\end{align}
where $a$ is the field, $a_0$ is a non-zero amplitude, $\omega$ is the excitation frequency, $\omega_0$ is the resonance frequency, and $\gamma$ is the damping parameter. From the Lorentzian distribution described in \eqref{eq:lorentzian}, we can readily get the stored energy bandwidth $\Delta \omega$, defined from the full width half maximum of the resonance, as $\Delta \omega=2\gamma$. From the temporal evolution of the resonance amplitude in \eqref{eq:timeevolve}, we can obtain the resonance lifetime ($\Delta T$), defined as the time it needs to reach $|a_0|e^{-1}$ from when the external excitation stops ($t=0$), as $\Delta T=1/\gamma$. Therefore, combining these two expression leads to the time bandwidth product of the single resonance condition, presented in Eq.~(6) of the main text \cite{mann2019nonreciprocal}.

In order to define the bandwidth limit of our meta-atom, we argue that the resonance lifetime can is identical to the group delay of the meta-atom. To support this argument, we first refer to the definition of Q-factor. For a general resonance problem, the Q-factor can be derived based on fractional bandwidth and frequency slope of the phase,
\begin{align}
Q=\frac{\omega_o}{\Delta \omega}, \quad  Q=\frac{\omega_0}{2}\bigg|\frac{\mathrm{d}\Phi}{\mathrm{d}\omega}\bigg|
\end{align}
From the Lorentzian resonance model and the resonance time-evolution described above, we can derive relationship between $\gamma$, $\Delta \omega$ and $Q$, as follow,
\begin{align}
\Delta \omega=\frac{\omega_o}{Q}=2\gamma \quad \Rightarrow \quad
\gamma = \frac{\omega_o}{2Q}
\end{align}
Therefore we have
\begin{align}
\Delta T=\frac{1}{\gamma} &= \frac{2Q}{\omega_o}= \bigg|\frac{\mathrm{d}\Phi_r}{\mathrm{d}\omega}\bigg|,
\end{align}
which confirms that the resonance lifetime is indeed the meta-atom group delay, i.e. the frequency slope of the phase $\Phi_r$. 

\section{\label{apx:derive_t0} Derivation of Minimum Additional Group Delay $t_0$}

The transition from negative to positive derivative of reactance marks the boundary between the Foster and non-Foster region, hence it determines the frequency range compatible with a passive, causal structure. It can be obtained by finding the zeros of the frequency derivative of $Z_{\mathrm{ms}}$
\begin{equation} \label{eq:z_ms_deriv_zero}
\frac{\mathrm{d}Z_{\mathrm{ms}}}{\mathrm{d\omega}}=0.
\end{equation}
Here, we apply this derivative operation to see the result in the dispersive metasurface case. The equation consists of four $\omega$ functions including two cotangent and tangent functions in the numerator and denominator. We notice that the derivative over frequency ($\omega$) is a quotient and product rule problem in which each of the implementing operation requires the use of chain rules. After replacing the cotangent with its tangent representation, we have,
\begin{equation}
\frac{\mathrm{d}Z_{\mathrm{ms}}}{\mathrm{d\omega}}=\frac{Z_o Z_s\big(\frac{2Z_o t_s}{\cos^2(\omega t_s)}-\frac{2Z_o t_s}{\sin^2(\frac{\pi x}{X}+\frac{\omega t_0}{2})\cos^2(\omega t_s)}-\frac{Z_s t_0}{\sin^2(\frac{\pi x}{X}+\frac{\omega t_0}{2})}+\frac{Z_s t_0}{\sin^2(\frac{\pi x}{X}+\frac{\omega t_0}{2})\cos^2(\omega t_s)}\big)\tan^2(\frac{\pi x}{X}+\frac{\omega t_0}{2})}{2(Z_o+Z_s\tan(\omega t_s)\tan(\frac{\pi x}{X}+\frac{\omega t_0}{2}))^2}
\end{equation}
Since we want to know the condition for $\frac{\mathrm{d}Z_{\mathrm{ms}}}{\mathrm{d\omega}}=0$, we seek the zeros of the numerator (having confirmed there is no contribution from poles on the denominator). After simplifying, we have
\begin{equation}
2Z_o t_s\sin^2\bigg(\frac{\pi x}{X}+\frac{\omega t_0}{2}\bigg)-2Z_o t_s-Z_s t_0\cos^2(\omega t_s)+Z_s t_0=0
\end{equation}
Applying the identity $\sin^2 x + \cos^2 x=1$, we have
\begin{equation}
\label{eq:Z_ms_zero_unsimplified_d}
-2Z_o t_s\cos^2\bigg(\frac{\pi x}{X}+\frac{\omega t_0}{2}\bigg)+Z_s t_0\sin^2(\omega t_s)=0.
\end{equation}
Simplifying Eq.~\eqref{eq:Z_ms_zero_unsimplified_d} leads to
\begin{equation}
\frac{2Z_o}{t_0} \cos^2\left(\frac{\pi x}{X}+\frac{\omega t_0}{2}\right)=\frac{Z_s}{t_s} \sin^2(\omega t_s)
\label{eq:zms_deriv_res_d}
\end{equation}

Inspecting Eq.~\eqref{eq:zms_deriv_res_d}, we note that the term $\cos^2\left(\frac{\pi x}{X}+\frac{\omega t_0}{2}\right)$ varies between 0 and 1 for any value of $\omega$. Therefore, we set this term to its maximum value of 1, which leads to the minimum additional group delay $t_0$ presented in the main text. We repeat the  calculation for the broadband achromatic case and confirmed that the minimum additional group delay $t_0$ also applies. The value of $t_0$ for the dispersive and achromatic case should be same if both are designed with the same substrate parameters and reflection angles. Therefore, the derived minimum group delay $t_0$, as presented in Eq.(10) of the main text, applies for both the achromatic and the dispersive metasurface case.

\begin{figure}[b]
	\subfloat{\includegraphics[width=0.23\linewidth]{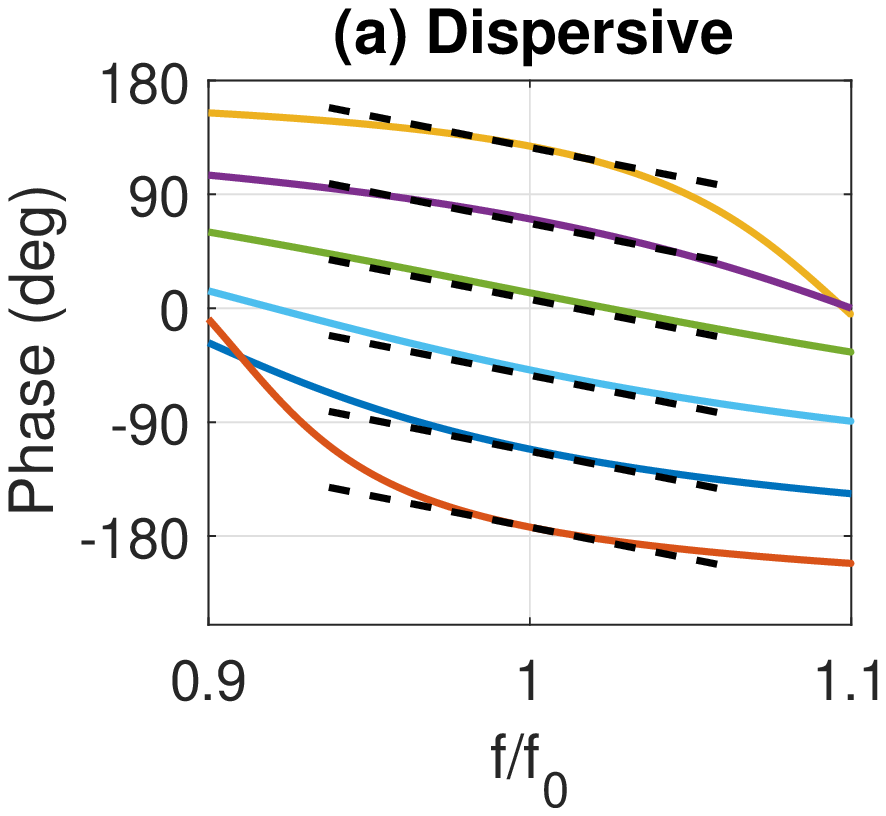}}
	\hfil
	\subfloat{\includegraphics[width=0.23\linewidth]{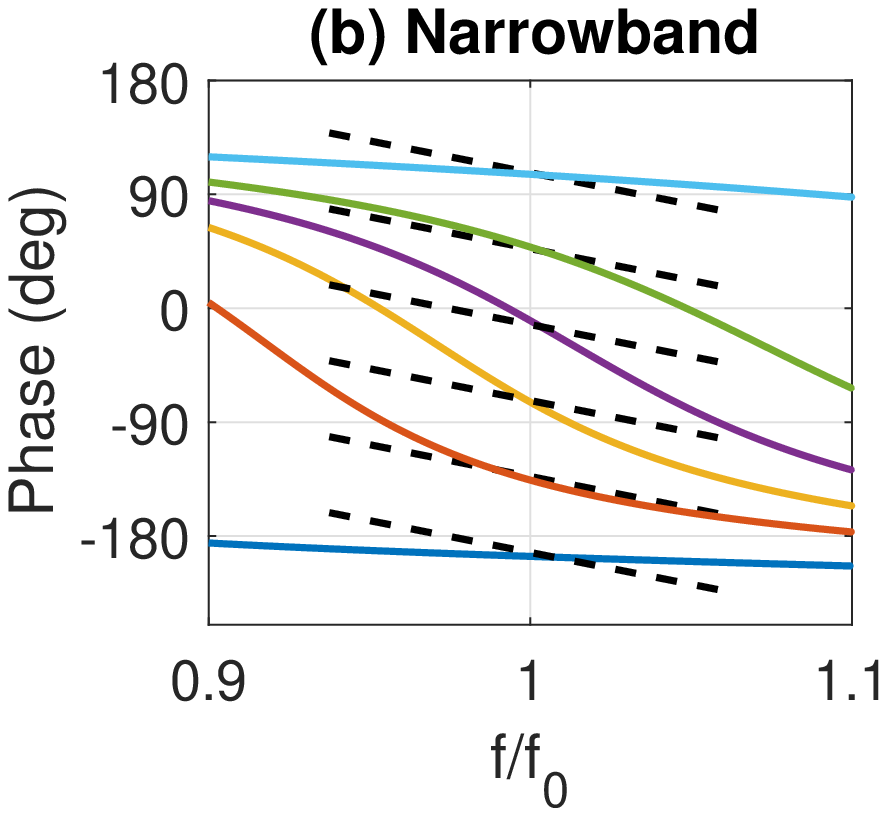}}
	\hfil
	\subfloat{\includegraphics[width=0.23\linewidth]{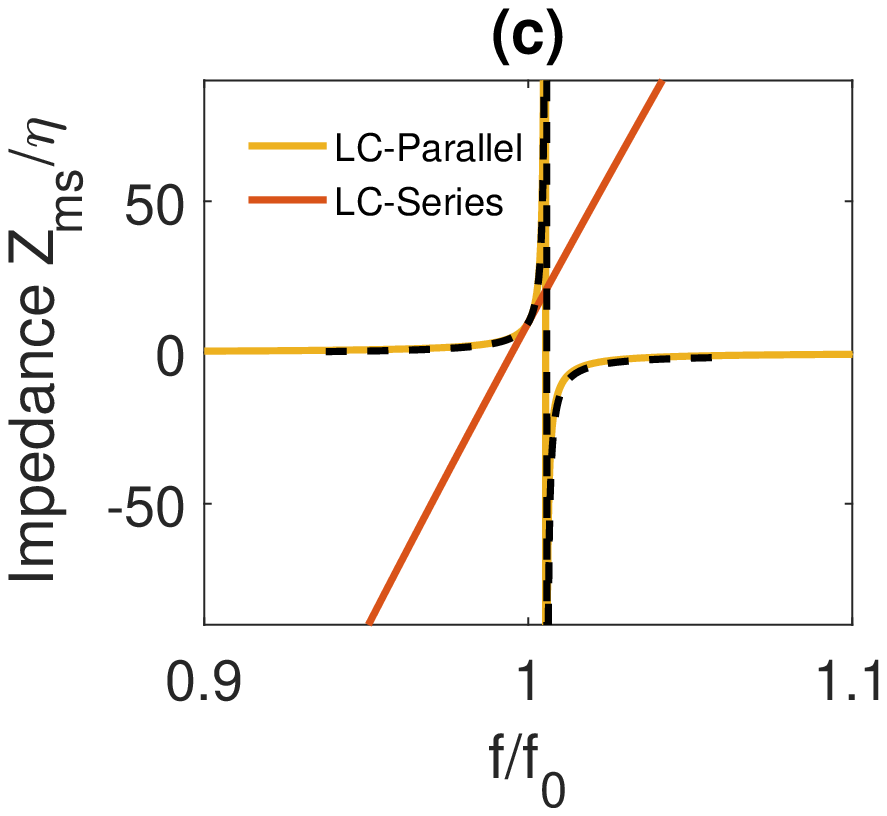}}
	\hfil
	\subfloat{\includegraphics[width=0.23\linewidth]{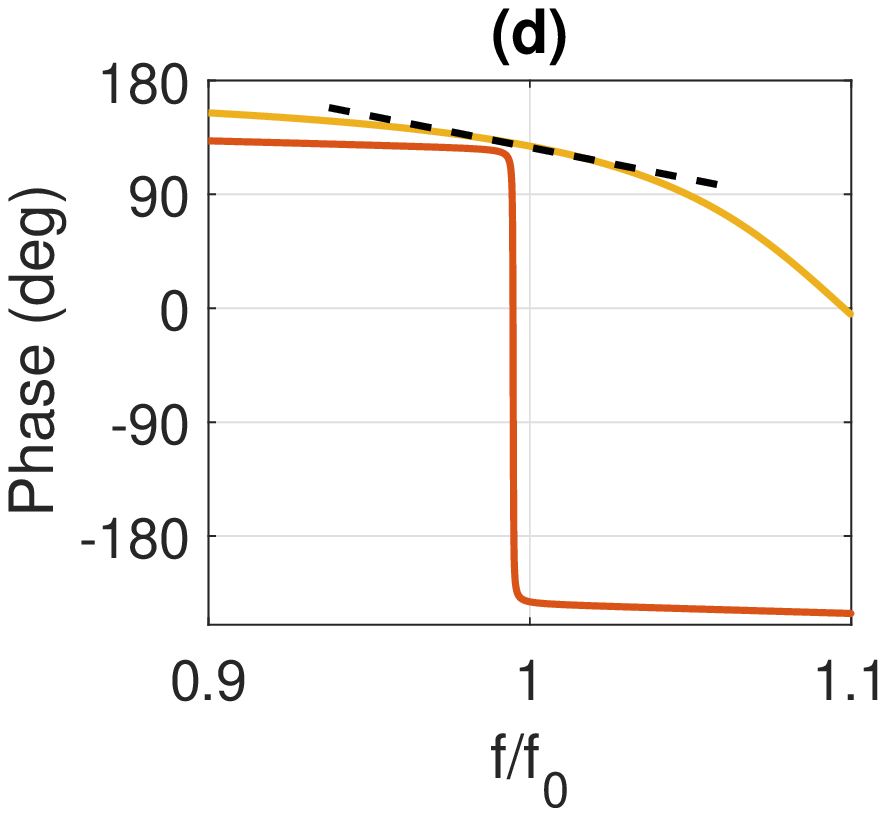}}
	\caption{\label{fig:phase_fit} Reflection phase for six unit cells of the broadband dispersive metasurface, (a) fitting both the impedance and its derivative and (b) fitting only the impedance (narrowband approach). The dashed lines show the target values, the solid lines show the fitted values. (c) Comparison of the series and parallel fits to the impedance for one cell of the metasurface, along with (d) the corresponding reflection phase.}
\end{figure}

\section{\label{apx:deciding_LC} LC-Series and Parallel Configuration for Meta-atoms}

To decide whether to use a series or parallel LC resonant equivalent circuit, we determine whether a pole or a zero is closest to the center frequency. An example of this is depicted in Fig.~\ref{fig:phase_fit}(c). Here, the pole is near the center frequency ($\mathrm{f_0}$), thus fitting to a series resonance, even with both impedance and its derivative considered, gives an incorrect broadband profile, as shown by the red curve. The reflection phase depicted in Fig.~\ref{fig:phase_fit}(d) shows a large discrepancy between the designed phase (dashed curve) and its LC-series fitting (red curve). A LC-parallel implementation gives better fitting to both the impedance and reflection phase as depicted by the yellow curves in Fig.~\ref{fig:phase_fit}(c) and (d). As shown in Fig.~\ref{fig:phase_fit}(a), by appropriately tailoring the series or parallel resonance according to the broadband dispersive design, the phase from the calculated transmission line model (continuous curve) matches the designed phase profile (dashed curve). However, the phase obtained by the narrowband approach as in Fig.~\ref{fig:phase_fit}(b), matches the design only at the center frequency.

\begin{figure}[t]
	\centering
	\includegraphics[width=0.49\linewidth]{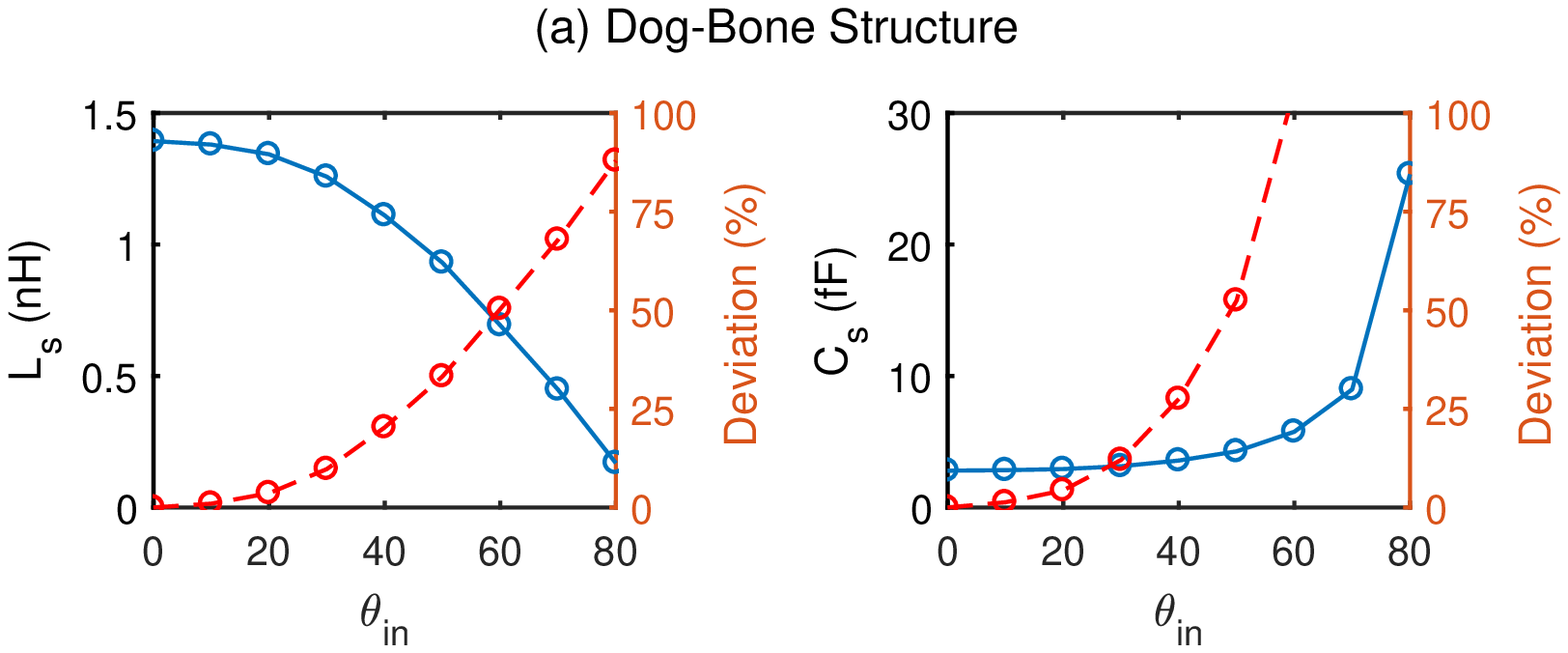} 
	\includegraphics[width=0.5\linewidth]{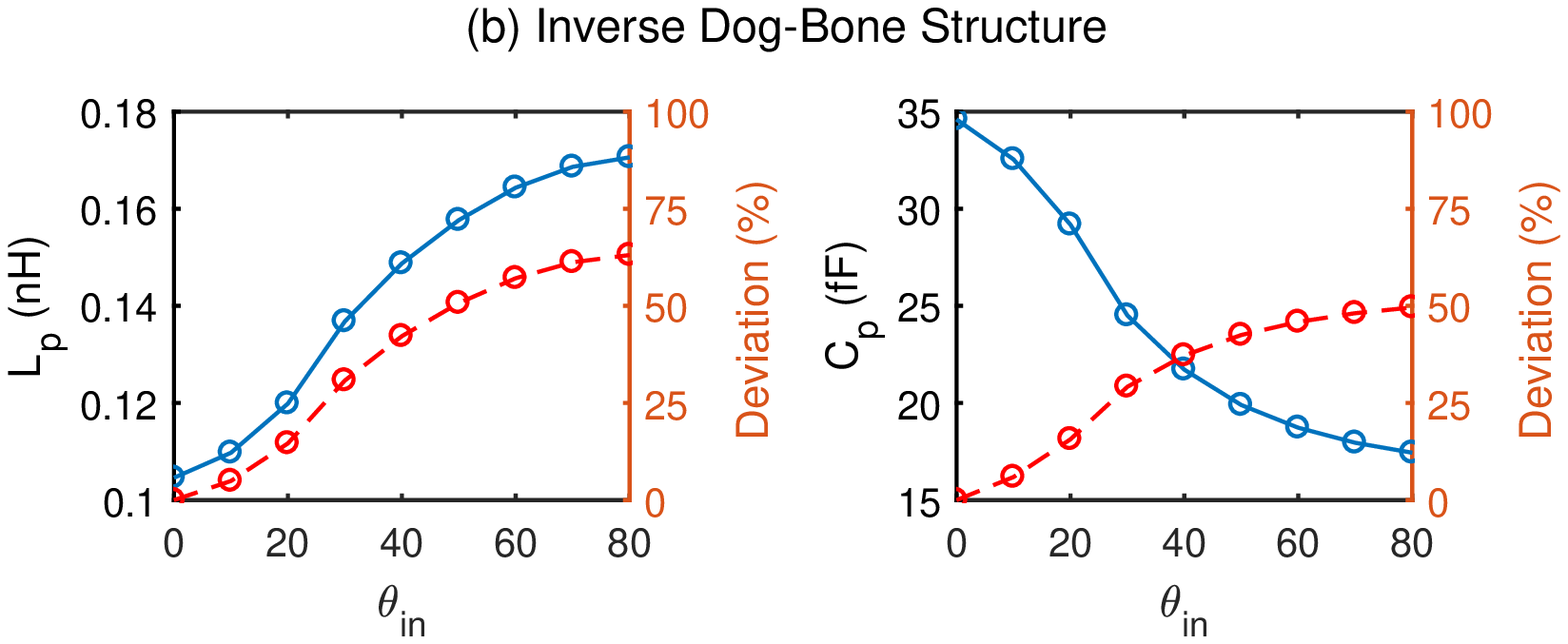}
	\caption{Results from L and C extractions with variation of the incident angles in (a) dog-bone structure (b) inverse dog-bone structure.}
	\label{fig:inc_cst}
\end{figure}

\section{\label{apx:lc_oblique} LC Extraction Under Oblique Incidence}
Given the non-local response of the meta-atom, spatial dispersion can be observed and oblique incidence can influence the extraction of LC parameters. A TE polarized wave incident upon the metasurface at non-normal angles will mainly contribute to the changes of local magnetic field. Therefore, for dog-bone structures, non-normal incidence has minimal effects, as they mainly interact with the electric field. For inverse dog-bone structures, non-normal angle of incidence has more pronounced effects and can contribute to significant changes in the values of $L_p$ and $C_p$. 

Fig.~\ref{fig:inc_cst} shows simulation results of changing the incidence angle ($\theta_{i}$) on the extracted L and C parameters from the dispersive metasurface design. In both extractions, the dependency of $t_s$ and $Z_{s}$ on the angle of incidence has been taken into account. We see that below 40$^\circ$, $L_s$ and $C_s$ deviates less than 25\% for the dog-bone structures while for the inverse dog-bone structures, $L_p$ and $C_p$  deviate almost 50\% from their values at normal incidence. This confirms that the parallel-LC configurations interact more strongly with the incoming magnetic fields.

\section{\label{apx:metaatom_para} Details of Meta-atom Realization}

The dog-bone and inverse dog-bone structures were simulated in CST, using a geometrical parameter sweep  to create a look-up table containing the extracted inductance and capacitance values. Based on the match to the required L and C, six geometrical combinations for the meta-atom implementation were chosen. Table~\ref{meta_atom_para_bb} shows details of the realized dog-bone and inverse dog-bone structures for the broadband dispersive metasurface. Here, unit cell no.~3 is implemented using an inverse dog-bone structure (parallel LC), while the rest are implemented using dog-bone structures (series LC). Note, that the initial look-up table shown by Fig.5(c, d) of the main text uses T=1.2mm and S=0.15mm. In several meta-atoms, the parameter T has been adjusted to account for the rounding effects (due to the etching resolution in fabrication).  

For the narrowband metasurface, the meta-atom implementation only requires one geometrical sweep for the dog-bone structure. The vertical size of the meta atom in this narrowband design is $D_y$=1.08\,mm, while for the broadband dispersive design it is $D_y$=2\,mm.

\begin{table}[h]
	\caption{\label{meta_atom_para_bb} Dog-bone and inverse dog-bone (indicated by asterisk) geometrical parameters (in millimeters) for the broadband dispersive and narrowband metasurface implementation. For the narrowband metasurface all meta-atoms are implementated as dog-bone structures.}
	\begin{ruledtabular}
		\begin{tabular}{|c | c c c c | c c c c |}
			{}& \multicolumn{4}{c|}{Broadband Dispersive} & \multicolumn{4}{c|}{Narrowband}\\
			{Cell No.} & {G}     & {W}     & T      & S   & {G}     & {W}     & T      & S   \\ \hline
			1          & 1.10    & 1.00    & 1.10   & 0.15 & 0.14     & 0.15    & 0.60   & 0.10\\
			2          & 1.32    & 0.36    & 1.00   & 0.15  & 0.40     & 0.15    & 0.60   & 0.10 \\
			3*         & 1.20    & 0.65    & 1.50   & 1.05  & 0.45     & 0.15    & 0.60   & 0.10 \\
			4          & 1.35    & 0.80    & 1.00   & 0.15 & 0.48     & 0.15    & 0.60   & 0.10 \\
			5          & 1.24    & 1.20    & 1.20   & 0.15  & 0.52     & 0.15    & 0.60   & 0.10 \\
			6          & 1.17    & 1.20    & 1.20   & 0.15 & 0.67     & 0.15    & 0.60   & 0.10
		\end{tabular}
	\end{ruledtabular}
\end{table}

\section{Details of Measurement Procedure}

The procedure for extracting Floquet harmonics involves four runs of far-field measurement, with four different incident angles as shown in Table \ref{floquet_table}. To account for the blind range of 50$^\circ$ due to the antennas that cannot overlap each other, non-normal incident angles were used. These incident angles were chosen to  ensure that the specular reflection and other Floquet harmonics were well captured by the measurement system. Additionally, Fabry-Perot resonances between the sample and exciting antenna were avoided by using these non-normal incident angles. As detailed in Sec.~\ref{apx:lc_oblique} of the Supplementary Material, the use of non-normal incidence below $30^\circ$ has very little effects on meta-atom L and C values, which should yield small deviation in the overall anomalous reflection performance.  

\begin{table}[h]
	\caption{\label{floquet_table} Different incident angles used to measure all possible Floquet harmonics from the metasurface}
		\begin{tabular}{p{0.2\linewidth}p{0.2\linewidth}p{0.2\linewidth}}
			\hline
			\multicolumn{1}{l}{Run} & \multicolumn{1}{l}{$\theta_{in}$} & \multicolumn{1}{l}{Floquet Mode} \\ \hline
			\mbox{1}                       & \mbox{-15}                                & \mbox{$\mathrm{F_0}$, $\mathrm{F_1}$}                                \\
			2                       & -5                                 & $\mathrm{F_2}$                                    \\
			3                       & 5                                  & $\mathrm{F_{-2}}$                                   \\
			4                       & 15                                 & $\mathrm{F_{-1}}$                                   \\ 
			\hline
		\end{tabular}
\end{table}


\end{document}